\documentclass[sigconf, authorversion]{acmart}

\AtBeginDocument{%
  \providecommand\BibTeX{{%
    \normalfont B\kern-0.5em{\scshape i\kern-0.25em b}\kern-0.8em\TeX}}}

\copyrightyear{2022}
\acmYear{2022}
\setcopyright{acmcopyright}
\acmConference[ASIA CCS '22] {Proceedings of the 2022 ACM Asia Conference on Computer and Communications Security}{May 30--June 3, 2022}{Nagasaki, Japan.}
\acmBooktitle{Proceedings of the 2022 ACM Asia Conference on Computer and Communications Security (ASIA CCS '22), May 30--June 3, 2022, Nagasaki, Japan}
\acmPrice{15.00}
\acmISBN{978-1-4503-9140-5/22/05}
\acmDOI{10.1145/3488932.3497766}

\usepackage{graphicx} 
\usepackage{comment} 
\usepackage{subfig} 

\graphicspath{ {./img/} }

\usepackage[ruled,vlined,linesnumbered]{algorithm2e}
\usepackage{listings}
\usepackage{multirow}

\usepackage{pifont}
\newcommand{\cmark}{\ding{51}}%
\newcommand{\xmark}{\ding{55}}%

\usepackage{enumitem}
\newcommand{\tinyskip}{\vspace{3pt}}
\newcommand{\mypar}[1]{\tinyskip\noindent\textbf{#1.}\xspace}

\usepackage{xcolor}

\newcommand*\yescheck[1]{%
  \expandafter\newcommand\csname #1yescheck\endcsname{\textcolor{#1}{\cmark}}%
}
\yescheck{blue}
\yescheck{green}
\yescheck{red}

\newcommand*\nocheck[1]{%
  \expandafter\newcommand\csname #1nocheck\endcsname{\textcolor{#1}{\xmark}}%
}
\nocheck{blue}
\nocheck{green}
\nocheck{red}

\newcommand*\maybecheck[1]{%
  \expandafter\newcommand\csname #1maybecheck\endcsname{\textcolor{#1}{$\sim$}}%
}
\maybecheck{yellow}
\maybecheck{green}
\maybecheck{red}

\definecolor{codegreen}{rgb}{0,0.6,0}
\definecolor{codegray}{rgb}{0.5,0.5,0.5}
\definecolor{codepurple}{rgb}{0.58,0,0.82}
\definecolor{backcolour}{rgb}{1,1,1}

\lstdefinestyle{mystyle}{
    backgroundcolor=\color{backcolour},   
    commentstyle=\color{codegreen},
    keywordstyle=\color{magenta},
    numberstyle=\tiny\color{codegray},
    stringstyle=\color{codepurple},
    basicstyle=\ttfamily\footnotesize,
    breakatwhitespace=false,         
    breaklines=false,                 
    captionpos=b,                    
    keepspaces=true,                 
    numbers=none,                    
    numbersep=5pt,                  
    showspaces=false,                
    showstringspaces=false,
    showtabs=false,                  
    tabsize=2
}
\long\def\authornote#1{%
        \leavevmode\unskip\raisebox{-3.5pt}{\rlap{$\scriptstyle\diamond$}}%
        \marginpar{\raggedright\hbadness=10000
        \def\baselinestretch{0.8}\tiny
        \it #1\par}}

\usepackage{color}
\usepackage{soul}
\definecolor{darkgreen}{rgb}{0,0.5,0}
\definecolor{purple}{rgb}{1,0,1}
\newcommand{\kibitz}[2]{\ifnum\Comments=1\textcolor{#1}{#2}\fi}

\newcount\Comments  
\newcommand{\attacknameLong}{\textsc{Alexa versus Alexa}}
\newcommand{\attackname}{\textsc{AvA}}

\usepackage{colortbl,dcolumn}
\definecolor{lightgreen}{rgb}{0.5,0.8,0}
\definecolor{darkgreen}{rgb}{0.2,0.7,0}
\definecolor{darkorange}{rgb}{0.66,0.35,0}

\usepackage{fontawesome}

\def\zz#1{%
\ifdim #1 pt > 8 pt \cellcolor{darkgreen} \else
\ifdim #1 pt > 6 pt \cellcolor{lightgreen}\else
\ifdim #1 pt > 4 pt \cellcolor{yellow}\else
\ifdim #1 pt > 2 pt \cellcolor{orange}\else
\cellcolor{red}\fi\fi\fi\fi
#1}

\usepackage{makecell}

\begin{document}
\fancyhead{}

\title{\attacknameLong{}: \\ Controlling Smart Speakers by Self-Issuing Voice Commands}


\author{Sergio Esposito}

\affiliation{%
  \institution{Royal Holloway, University of London}
  \country{United Kingdom}
}
\email{sergio.esposito.2019@live.rhul.ac.uk}

\author{Daniele Sgandurra}
\affiliation{%
  \institution{Royal Holloway, University of London}
  \country{United Kingdom}
}
\email{daniele.sgandurra@rhul.ac.uk}

\author{Giampaolo Bella}
\affiliation{%
  \institution{Università degli Studi di Catania}
  \country{Italy}
}
\email{giamp@dmi.unict.it}

\renewcommand{\shortauthors}{Esposito, et al.}

\begin{abstract}
We present \attacknameLong{} (\attackname{}), a novel attack that leverages audio files containing voice commands and audio reproduction methods in an offensive fashion, to gain control of Amazon Echo devices for a prolonged amount of time. \attackname{} leverages the fact that Alexa running on an Echo device correctly interprets voice commands originated from audio files even when they are played by the device itself -- i.e., it leverages a \textit{command self-issue} vulnerability. Hence, \attackname{} removes the necessity of having a rogue speaker in proximity of the victim's Echo, a constraint that many attacks share. With \attackname{}, an attacker can self-issue any permissible command to Echo, controlling it on behalf of the legitimate user. We have verified that, via \attackname{}, attackers can control smart appliances within the household, buy unwanted items, tamper linked calendars and eavesdrop on the user.
We also discovered two additional Echo vulnerabilities, which we call \textit{Full Volume} and \textit{\texttt{Break} Tag Chain}. The \textit{Full Volume} increases the self-issue command recognition rate, by doubling it on average, hence allowing attackers to perform additional self-issue commands. \textit{\texttt{Break} Tag Chain} increases the time a skill can run without user interaction, from eight seconds to more than one hour, hence enabling attackers to setup realistic social engineering scenarios. By exploiting these vulnerabilities, the adversary can self-issue commands that are correctly executed 99\% of the times and can keep control of the device for a prolonged amount of time. 
We reported these vulnerabilities to Amazon via their vulnerability research program, who rated them with a Medium severity score. In addition, we discuss the results of a set of tests performed on three voluntary Echo-equipped households to verify the feasibility of \attackname{} in real scenarios, finding that the attack remains undetected and operative in most cases. Finally, to assess limitations of \attackname{} on a larger scale, we provide the results of a survey performed on a study group of 18 users, and we show that most of the limitations against \attackname{} are hardly used in practice.
\end{abstract}






\maketitle


\section{Introduction}

An increasing number of households and companies are making use of IoT devices through their daily activities. In 2019, there were more than 143 million ``Smart Homes'' worldwide, and their number is growing at a steady pace~\cite{statista1}. Smart devices feature security commodities as cameras and baby monitors, entertainment-ware such as TVs, and much more. Voice Personal Assistants (VPA) and smart speakers, such as Amazon Echo and Google Home, were in use in approximately 35\% of US households in 2019, with an expected increase to 75\% by 2025~\cite{statista2}. VPAs are defined as \textit{``software agents that can interpret human speech and respond via synthesized voices''} \cite{hoyVPAdefinition}. Smart speakers, in fact, feature a microphone to capture voice commands, along with a speaker to answer user queries.

As consumer-grade VPAs are recent developments, research has begun to explore new angles for exploitation of such devices. For example, hidden commands are audio tracks containing voice commands generated via Adversarial Machine Learning, in such a way that an Automatic Speech Recognition (ASR) system misclassifies their content: the human ear can hear a sentence \textit{s} (or even no sentence at all~\cite{dolphinattack}), which is instead classified by the ASR as another sentence \textit{s'}, chosen by the attacker. Some works assume a black-box scenario \cite{taori, sirenattack, didyouhearthat, kenansville}, where the attacker does not know the internal mechanisms of the ASR system; other works operate with white-box assumptions \cite{houdini, carliniaudioadv, psychoacoustichiding, qin, abdoli}, where the attacker has access to the ASR's models and source code. Some works succeed even over-the-air \cite{devilwhisper, commandersong, hiddenvoicecommands, imperio, Yakura, cocainenoodles, practicalhvc}, that is when the command is not directly fed into the ASR system, but it is emitted from a speaker, it travels over the air and then it gets captured by a microphone. Adversarial attacks can also have different goals, for example hindering command recognition \cite{adversarialmusic} or bypassing security measures \cite{whoisrealbob, kreuk}. Other attacks against smart speakers deceive the user into thinking they are talking with the VPA or with a legitimate application, when they are actually talking with an attacker-controlled application or device~\cite{skillsquatting, lyexa, dangerousskills}. However, most of these works \cite{lyexa, devilwhisper, adversarialmusic} leverage external speakers to issue commands to a VPA, reducing the overall likelihood of the attack.

Our attack, ``\attacknameLong{}'' (\attackname{}), is the first to exploit the vulnerability of self-issuing arbitrary commands on Echo devices, allowing an attacker to control them for a prolonged amount of time. With this work, we remove the necessity of having an external speaker near the target device, increasing the overall likelihood of the attack. \attackname{} starts when the Echo device begins streaming an audio file that contains voice commands. This can be done, for example, by opening a malicious \emph{skill}\footnote{Henceforth, we will always refer to applications for VPA as \textit{skills}, which is the name used in the Amazon Alexa's context. For the sake of brevity, we sometimes say that the skills ``run on the Echo device'', however, skills actually run on their cloud hosting.} that makes the Echo device tune in a radio station that streams such voice commands: they are emitted from Echo's speakers and captured by Echo's microphone, and are effectively self-issued. Since the commands streamed by the radio station can be altered by the attacker on-the-fly, such radio station results in a Command \& Control (C\&C) server. We show that, with \attackname{}, the attacker can make the Echo device perform any permissible action, such as controlling smart appliances in the victim's household (e.g., lights and door-locks), calling any phone number or starting other skills.

We also illustrate how, as a post-exploitation action, the adversary can leverage \attackname{} to open another malicious skill, which is able to gather user commands and spoof other skills' behaviour~\cite{dangerousskills}. This allows the attacker to perform personal data theft and social engineering actions. Furthermore, we demonstrate that the adversary is able to keep the malicious skill running for a prolonged amount of time, independently from user interaction.

\textbf{Contributions.} The paper makes these contributions:

\begin{itemize}

\item It presents \attacknameLong{} (\attackname{)}, an attack that leverages the Echo's self-issue vulnerability (\S\ref{exploitation}); we evaluate its effectiveness with multiple payloads (\S\ref{payloadsection}), scenarios (\S\ref{evaluation}) and attack vectors (\S\ref{attackvectors}).

\item It describes two vulnerabilities we found during the tests: the \textit{Full Volume} vulnerability (\S\ref{sec:fvv}) allows attackers to self-issue any command without any reduction of the device volume, while the \texttt{Break} \textit{Tag Chain} vulnerability (\S\ref{persistency}) allows attackers to violate Amazon's SSML (Speech Synthesis Markup Language) policy.

\item It evaluates feasibility and impact of \attackname{}, both alone and when used in combination with the two vulnerabilities (\S\ref{sec:Feasibility}).

\item It details the results of a field study of \attackname{} within three voluntary households (\S\ref{userstudy}), and of a survey submitted to a study group composed of 18 Amazon Echo users (\S\ref{limitations}): we show the attack is feasible and limitations are mostly theoretical.
\end{itemize}

\section{Research Goals and Threat Model}
\label{goals}

The main research questions we address in this work are: \emph{``Is the self-issue vulnerability a real threat? To what extent does it allow an attacker to control, in a real scenario, an Echo device without leveraging any external speaker? What impact does it have?''} From the above questions stem three distinct research goals (RG):

\begin{itemize}
\item \textbf{RG1: Feasibility.} We want to investigate if it is possible to exploit the self-issue vulnerability in real scenarios (\S\ref{evaluation}), and to assess its limitations (\S\ref{limitations}).

\item \textbf{RG2: Impact.} We want to understand whether being able to self-issue commands can lead to critical scenarios that are detrimental for the user's security and privacy (\S\ref{impactsection}).

\item \textbf{RG3: Countermeasures.} We want to understand what countermeasures could the users or the device manufacturers apply to defend against such threat (\S\ref{countermeasures}).

\end{itemize}

\subsection{Threat Model}
\label{threatmodel}

\attackname{} is the offensive act of self-issuing arbitrary commands on an Echo device. Using \attackname{}, an attacker can control victim Echo devices leveraging common audio reproduction methods, such as a radio station that acts like a C\&C server, or making Echo Dot act as a speaker for a nearby device. As a post-exploitation action, the attacker can set up a Voice Masquerading Attack (VMA)~\cite{dangerousskills} to gather user commands and reply to them arbitrarily.

\begin{figure}[!hbt]
    \centering
    \includegraphics[scale=0.43]{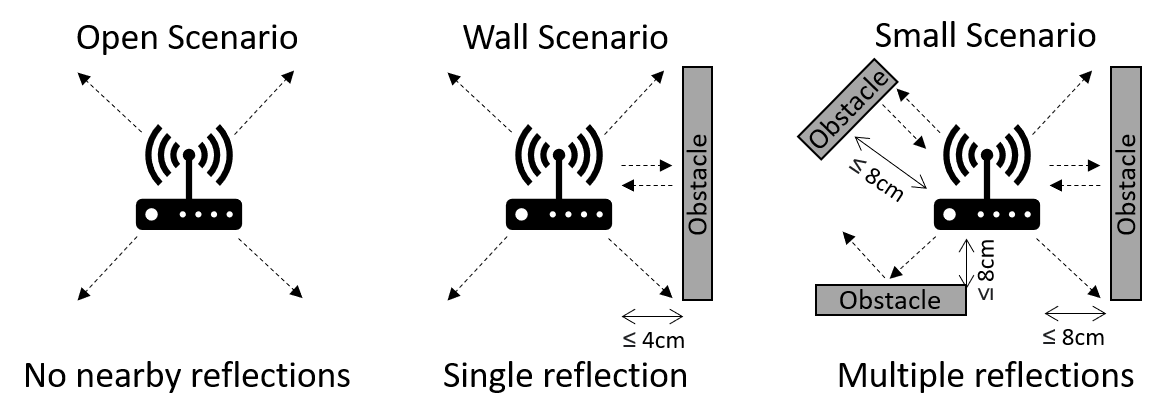}
    \caption{Soundwave Reflection in the Different Scenarios}
    \label{fig:scenarios}
\end{figure}

We assume the attacker does not have any knowledge of the ASR system's internal mechanisms, and that people in the household of the victim Echo will interact with it from time to time. The target Echo can be placed anywhere inside the house. As soundwaves emitted by Echo are reflected differently if there are obstacles nearby, we consider three scenarios (also depicted in Fig.~\ref{fig:scenarios}): (i) \textbf{Open Scenario}: there are no obstacles near Echo, as it happens on a conference table; (ii) \textbf{Wall Scenario}: Echo is placed near a wall, and distance from the wall is approximately 1.5cm to 4cm, and the closest obstacle is farther than 8cm; (iii) \textbf{Small Scenario}: Echo is placed on a surface with other objects on it (the wall can count as an object), and distance from at least 2 obstacles must be 1.5cm to 8cm.

If the attacker is near the target Echo, they can exploit the self-issue via a Bluetooth device, e.g. their smartphone: note that this operation does not require any PIN (hence, no bruteforce or other similar attacks are required), and the pairing requires approximately 25 seconds. Additionally, once paired, the Bluetooth device can connect and disconnect from Echo without any need to perform the pairing process again. Therefore, the actual attack may happen several days after the pairing (assuming the attacker uses the same paired Bluetooth device). If the attacker is not near the target Echo, they can use a radio station skill to self-issue commands: in this case, we assume the attacker has already deployed such skill on the Alexa Skill Store, so it will be immediately available. We also assume that the attacker has deployed another skill that performs a Voice Masquerading Attack, which we call \textit{Mask Attack}: this skill allows the attacker to listen to user commands.  No other equipment (e.g., external speakers) is needed for \attackname{}.

\mypar{Publishing a Malicious Skill}
Currently, anyone can deploy skills on the store, and skills do not need any special permission to run on the device or to play audio. Note that skills do not need to be installed, e.g. like apps on smartphones. There are documented cases of policy-violating skills successfully passing the validation \cite{dangerousskillsgotcertified}. Once a skill passes the certification process, further modifications to its code can be deployed to the live skill without having to re-certify it again \cite{areyouhomealone, dangerousskillsgotcertified}, since the code resides on a server external to AVS. Hence, an attacker can certify a harmless skill and add malicious capabilities afterwards.

\subsection{Attack's Success}
We assume the \attackname{} attack to be successful if the adversary manages to achieve any of these goals: (i) \textbf{Undermine the physical safety of the user}, for example by opening the front door or turning on the heaters; (ii) \textbf{Violate the user's privacy}, e.g. by disclosing personal or sensitive data, passwords or PINs; (iii) \textbf{Perform malicious actions}, for example, buying items on Amazon on behalf of the user. Feasibility and limitations of \attackname{} under this threat model are discussed, respectively, in \S\ref{sec:Feasibility} and \S\ref{limitations}.

\section{Attack Preparation \& Exploitation}\label{sec:prep}
We now describe the preparation and exploitation steps  for \attackname{}.

\subsection{Attack Vectors: Playing Audio Files}
\label{attackvectors}
It all starts by investigating the possible ways to play audio files on an Echo device. We want to find out whether they can be used in an offensive fashion to bootstrap \attackname{}, that is, if they can be used as attack vectors. There are three ways to play audio on an Echo device: (i) \textbf{Radio Station}: the Echo device tunes in a radio station. This can be done by means of the Music and Radio skills; (ii) \textbf{Bluetooth Audio Streaming}: another device, e.g. a PC or a smartphone, connects to Echo via Bluetooth and streams audio on Echo's speaker; (iii) \textbf{SSML \texttt{audio} Tag}: a skill that contains an \texttt{audio} Speech Synthesis Markup Language (SSML)~\cite{ssml} tag is opened, and the audio file specified by such SSML tag is played by Echo.

However, not all these methods are suitable for use with \attackname{}. In fact, we found that, to be able to self-issue a command to Echo, a property, which we call \emph{non-exclusivity of the audio channel}, must be true. In fact, for the Echo device to be able to capture the whole audio command, the streaming must not be stopped when the Echo device hears the wake-word (usually \textit{Alexa} or \textit{Echo}). Table~\ref{tab:attackvectors} shows the valid attack vectors we found and their peculiarities (see Appendix~\ref{app:vectors} for more details). In particular, the Radio Station works remotely and can be used to control multiple devices at once, which allows the attacker to reach more targets than the Bluetooth vector. However, the adversary is not able to use the Full Volume Vulnerability (FVV) and they need Social Engineering to make the user start the radio station. Furthermore, Music \& Radio skills cannot be developed outside the US. Additionally, if the user closes the radio station, the adversary would have to go over the whole process again to reconnect to the target Echo. By contrast, the Bluetooth vector works locally and with one device at the time, but it will not encounter all the other limitations. In particular, once the Bluetooth pairing between the target Echo and the adversary's device takes place, even if the latter is disconnected, they can always reconnect in a second moment without repeating the pairing process.

\begin{table}[!h]
    \caption{Attack Vectors Summary}
    \label{tab:attackvectors}
    \begin{center}
    \resizebox{\columnwidth}{!}{
    \begin{tabular}{ l c c c c c c }
        \Xhline{2\arrayrulewidth}
        \textbf{Atk Vector} & \textbf{Remote} & \textbf{Multiple} & \textbf{FVV} & \textbf{Worldwide} & \textbf{SE Not Needed} & \textbf{Can Restart} \\
        \hline

        Radio Station & \greenyescheck & \greenyescheck & \rednocheck  & \rednocheck  & \rednocheck & \rednocheck  \\
        Bluetooth & \rednocheck & \rednocheck & \greenyescheck  & \greenyescheck  & \greenyescheck & \greenyescheck \\
        
        \Xhline{2\arrayrulewidth}
    \end{tabular}
    }
    \end{center}
    \footnotesize{\textbf{Remote:} Works remotely | \textbf{Multiple:} Can control multiple Echo devices at once | \textbf{FVV:} Can be used with the Full Volume Vulnerability | \textbf{Worldwide:} Attack Vector is available anywhere in the world | \textbf{SE Not Needed:} Adversary does not need Social Engineering to start the attack | \textbf{Can Restart:} If connection to the attack vector terminates, the adversary can reconnect without going through the initial steps.}
\end{table}

\subsection{Payloads: Generating Audio Files}

\label{payloadsection}
In \attackname{}, a payload is the malicious audio containing a wakeword and the attacker's command. We identify three possible payload types: (i) \textbf{Text-To-Speech (TTS) Voice Commands}, generated via any TTS solution; (ii) \textbf{Hidden Commands}, generated via solutions that output adversarial samples working against Alexa over-the-air; (iii) \textbf{Real-Voice Commands}, recorded by the attacker with their own voice or with someone else's.

With regards to the first payload type, henceforth we refer to Google TTS to generate malicious audio commands and evaluate \attackname{}. This is arbitrary and the attacker can choose any TTS solution. For what concerns the second payload type, we performed extensive tests with state-of-the-art tools for the generation of adversarial commands that work over-the-air: some adversarial commands were successfully self-issued, however the success rate was rather low, hence an attacker cannot reliably use them for a real attack (we report our findings in Appendix~\ref{attackwithhidden}). Finally, regarding the third payload type, we realistically argue that the adversary would rather not use their own voice for the attack, and recording other people while issuing commands can be very impractical. Hence, the third payload type is not ideal for \attackname{} as well, and we will focus on the evaluation of TTS commands in Section~\ref{evaluation}. 

\subsection{Exploitation: Echo Self-Activation}
\label{exploitation}

\begin{figure}[!hbt]
    \centering
    \includegraphics[scale=0.3]{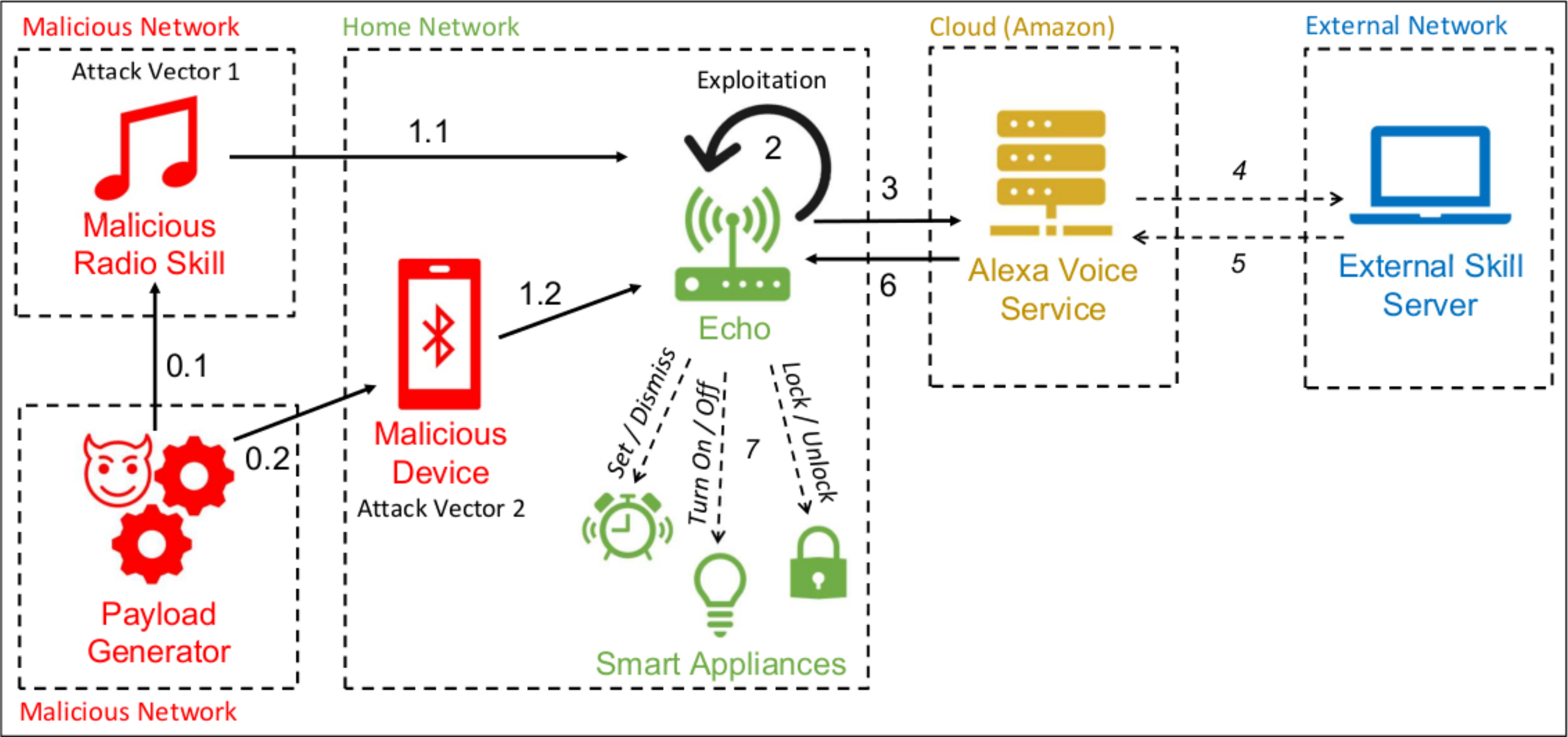}
    \caption{\attackname{}'s Exploitation Flow: Command Generation, Attack Vectors and Exploitation}
    \label{explimg}
\end{figure}

As discussed earlier, the self-activation of the Echo device happens when an audio file reproduced by the device itself contains a voice command. Regardless of the chosen attack vector, the adversary can freely self-issue any voice command for a prolonged amount of time. Fig.~\ref{explimg} shows the exploitation flow: the adversary sends a command (steps $0.x$ and $1.x$, being $x$ attack vector number), which is self-issued by the Echo device (step 2) and interpreted by AVS (step 3). If an external skill is requested by the command, AVS communicates with the related server (steps 4 and 5, indicated by a dotted line since this is an optional action), then it sends back the reply to Echo (step 6). As a result, the attacker can perform any action on the VPA (e.g. make phone calls, set alarms), on any skill (e.g., buy items), or they can control other smart appliances in the household (e.g., lights and door-locks) (step 7). 

\mypar{Full Volume Vulnerability}\label{sec:fvv}
During our tests using the Bluetooth attack vector, we noticed that sometimes the self-issued commands are played at full volume even after the wakeword recognition. Upon further inspection, we managed to reproduce this behaviour by self-issuing the command ``Echo, turn off''. In fact, subsequently, the affected Echo device does not turn down the volume anymore for the whole duration of the current audio stream, allowing the attacker to self-issue commands at the current volume in their entirety. We call this the \textit{Full Volume Vulnerability} (FVV).

We believe this is due to the fact that, when Echo is being used as a Bluetooth speaker and the ``turn off'' command is received, the audio stream should be stopped (as it happens with Music and Radio skills), but it does not. Hence, when another command is received, Echo does not turn down the volume because it assumes the reproduction has already ended. FVV has a great impact on the success rate of the self-issued commands, which is evaluated in Section~\ref{evaluation}. We reported this vulnerability to Amazon (see Appendix~\ref{responsibledisclosure}).

\section{Post-Exploitation Actions with Mask Attack}

\label{postexploitation}
\begin{figure}[!hbt]
    \centering
    \includegraphics[scale=0.55]{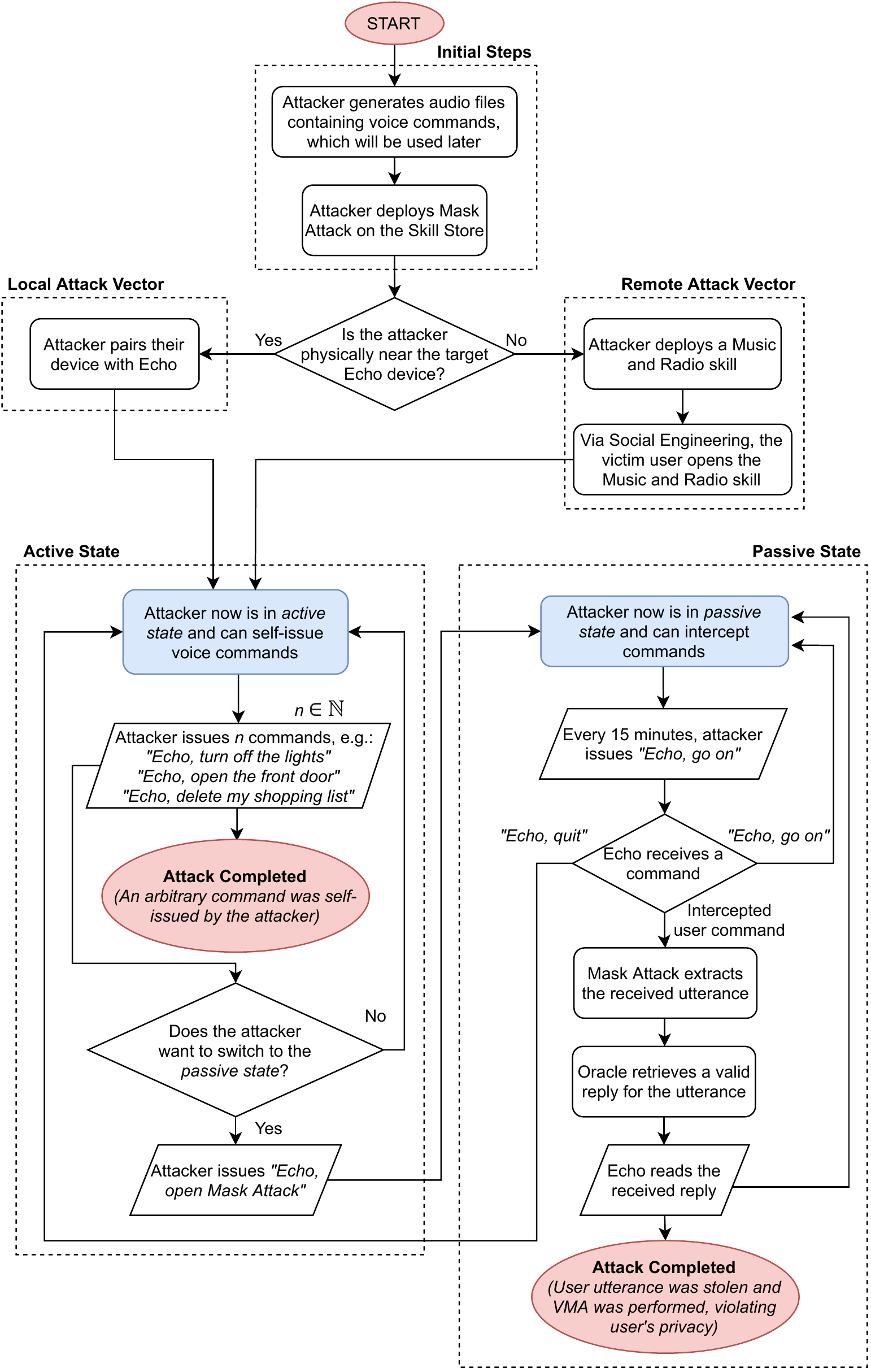}
    \caption{Process Flow Diagram for AvA}
    \label{fullFlow}
\end{figure}

In Figure~\ref{fullFlow} we summarize the main attack flow for \attackname{}. In particular, after the exploitation phase, the adversary can switch from the \textit{active state}, i.e. of self-issuing commands, to a \textit{passive state}, in which attackers can listen to commands issued by the user and capture personal data that could be sent along with the voice commands, for example the victim's home address. To this aim, the adversary exploits the self-issue vulnerability to open the \emph{Mask Attack} skill, which performs our original instance of the Voice Masquerading Attack~\cite{dangerousskills, lyexa} able to gather user commands and to issue valid replies for them. Such skill can trick the user into thinking that they are talking with the VPA or with any skill -- it also allows the adversary to arbitrarily answer any user query, for example by faking a PIN request to steal it. In fact, some sensitive actions require the user to say a PIN as a security measure. Additionally, by combining the self-issue with the SSML \texttt{break} tag chain in a skill's response, we found that it is possible to keep the Mask Attack skill running on the Echo device for a prolonged amount of time, even when the user does not interact with such skill. Table~\ref{tab:stateexample} gives some examples of actions that can be performed by the adversary in the active and passive states.

\subsection{Keeping the Mask Attack Skill Open}
\label{persistency}
Most skills follow a standard flow:\footnote{Some special skills, like Music and Radio skills, do not follow this flow.} when they are started, Alexa reads some text and then it prompts the user to reply, to continue the interaction. The user then has 8 seconds to do so, otherwise the skill is closed. However, if the user (or a self-issued command) interrupts Alexa while she is speaking by saying the wakeword and issuing a command, the VPA will try to execute this command in the skill's context. This is the key point of \attackname{}'s VMA approach.

In fact, Mask Attack appends several \texttt{break} SSML tags to every response: when read by Alexa, such tags add a pause of customizable length in her speech. We discovered that it is possible to violate Amazon's policy regarding SSML \texttt{break} tags, which states that \textit{``\texttt{break} tag silence cannot exceed 10 seconds, including scenarios with consecutive break tags. SSML with more than 10 seconds of silence isn't rendered to the user''}~\cite{ssml}. More specifically, we found that the only limit for this chain of \texttt{break} tags is the maximum length allowed for the \texttt{outputSpeech} property of a skill response, set by Amazon to 8000 characters~\cite{jsonreference}.

This allows Mask Attack to append more than 400 \texttt{break} tags, of ten seconds each, to every reply, adding up to over one hour of silence. Hence, every time that the device receives a command within this timespan, such command will be executed in the skill's context, refreshing the timeout timer. We do not need the user to issue a command every hour: through \attackname{}, we can self-issue the ``Echo, go on'' command, which activates an Intent that just outputs another hour of \texttt{break} tags. We call this Intent the \textit{ContinueIntent}. Additionally, we observed that the mere activation of the device with the wakeword is sufficient to refresh the skill timeout, as the last executed Intent is triggered again. The attacker can return to the active state by self-issuing the ``Echo, quit'' command. Any other command self-issued while in the passive state would be intercepted by Mask Attack, including ones that include skill names.

\begin{table}[hbt!]
    \caption{Examples of Malicious Actions to Perform in the Active and Passive States}
    \label{tab:stateexample}
    \footnotesize
    \begin{center}
    \begin{tabular}{ l l }
        
        \Xhline{2\arrayrulewidth}
        \textbf{Active State} & \textbf{Passive State} \\
        \textit{(Give Commands)} & \textit{(Intercept Commands / VMA)} \\
        \hline
        
        Control smart appliances & Capture commands \\
        
        \rowcolor{gray!25}
        Tamper linked calendars & Ask for personal data \\
        
        Reply/delete emails* & Ask for passwords/PINs \\
        
        \rowcolor{gray!25}
        Call any phone number* & Tamper replies \\
        
        Buy items on Amazon & Infer user habits \\
        
        \Xhline{2\arrayrulewidth}

    \end{tabular}
    
    \textbf{*}This feature of Alexa is available only in certain countries.
    \end{center}
\end{table}

\subsection{Retrieving Utterances \& Realistic Replies}
\label{subStealth}
The Voice Masquerading Attack (VMA) needs to retrieve a realistic response for any user command, and we adopt a solution that is similar to the one adopted in Lyexa ~\cite{lyexa}. However, since \attackname{} uses only the victim device, \attackname{} does not need to timely inject the ``use Mask Attack'' command to every user query, because the user is already talking with that skill. Our VMA solution consists of two components that communicate via custom APIs: (i) the Mask Attack skill; and (ii) ``The Oracle'', an external script that runs on another server controlled by the attacker. When the user, unknowingly, says a command to Mask Attack, the malicious skill has to retrieve the user utterance, since AVS does not expose it directly. To this end, we train a custom Slot with some alphanumerical dummy values, so that any sentence can fit the Slot, and then we introduce a new custom Intent within the skill, whose only sample utterance is the Slot itself. We call this intent \textit{InterceptIntent}. Since the only other Intent, \textit{ContinueIntent} (described in \S~\ref{persistency}) only activates itself with the ``Echo, go on'' command, almost every other utterance\footnote{Some utterances are reserved and they always trigger standard actions. For example, ``close'', ``stop'', ``exit'', and ``quit'' always close the active skill.} will activate \textit{InterceptIntent}. This allows Mask Attack to retrieve the content of its Slot, that is, the utterance. Note that \textit{InterceptIntent} activates even when the user tries to call another skill (e.g. ``Echo, ask Big Sky what's the weather like in Paris'').

The Mask Attack skill then needs to retrieve a realistic response for the user query. To this end, the Mask Attack skill sends the user utterance to the Oracle, which leverages the Alexa Voice Service (AVS) Client package~\cite{alexaclient} to asynchronously query AVS to obtain the real reply. To communicate with AVS, the Oracle transforms the plaintext utterance into an audio file using Google TTS.\footnote{Any other online Text-to-Speech service would work, as long as it can output \texttt{LINEAR16}~\cite{audioencoding} encoded audio files, which is the format accepted by AVS.} The Oracle then receives the reply to the query from AVS, within one or more mp3 audio files. The Oracle transforms the reply in text using Google Speech-To-Text (STT) and sends it back to Mask Attack. Mask Attack now has a realistic reply for the user's query and reads it to the user. The whole process takes approximately 5 seconds.

\section{Implementation and Evaluation}
Here, we first discuss the implementation of \attackname{}, and then we evaluate \attackname{} itself.

\subsection{Implementation}
\label{equip}

\mypar{Payload Generation and Exploitation}
We used Google TTS to generate voice commands. For the exploitation process, we streamed all payloads over Echo using two standard laptops running Windows 10 Pro 64-bit and Ubuntu 20.04, connecting them to Echo via Bluetooth (SBC codec).

\mypar{Post-Exploitation}
The post-exploitation of \attackname{} is composed of two modules: the Mask Attack skill and the Oracle. Mask Attack is written in Node.js and is hosted on Amazon Lambda. It uses the \texttt{ask-sdk-core} and \texttt{axios} packages. We also used \path{cookbook/progressive-response/v1} as CustomUserAgent. The Oracle is written in Python 3 and leverages the Google Cloud \texttt{speech} and \texttt{texttospeech} packages, along with \texttt{AlexaClient}. The Oracle also connects to a set of APIs we coded in PHP to communicate with a MySQL database, where we store received user utterances and their replies.

\subsection{Evaluation}
\label{evaluation}
We evaluate \attackname{} against a 3rd Generation Echo Dot. Since Music and Radio skills are available only in US, we focus on the Bluetooth attack vector (\S\ref{attackvectors}) to evaluate an attack scenario that is feasible anywhere. All payloads were streamed to the Echo Dot via the Bluetooth SBC codec. We also verified that both the self-issue vulnerability and the Full Volume Vulnerability can be successfully used on 4th Generation Echo Dot devices, although the success rates would differ from the ones reported in this paper, since this evaluation was performed on a 3rd Generation device.

\mypar{TTS Payload Performance}
\label{attackwithplain}
In the first test, we generated 70 audio payloads using Google TTS, and measured their effectiveness when self-issued to Echo. The samples include 7 commands, each generated with 10 different Google TTS voice profiles, namely from ``en-US-Wavenet-A'' to ``en-US-Wavenet-J''. In all the tests, the volume of the Echo Dot was set to 5 (out of 10) and the language was set to ``English (United States)''. Recall the definition of the three possible placements of the Echo device, given in Section~\ref{threatmodel}.

\begin{table}[hbt!]
    \caption{Self-Issued TTS Commands Reliability at Volume 5}
    \label{pvctable}
    \centering
    \resizebox{\columnwidth}{!}{%
    \begin{tabular}{ |c|c c c|c c c|c c c| }
        \hline
        \multirow{3}{*}{\textbf{TTS Voice Command}} & \multicolumn{9}{|c|}{\textbf{Google TTS Voice Profile and Scenario}} \\ \cline{2-10}
        & \multicolumn{3}{|c|}{\textbf{en-US-Wavenet-A}} &  \multicolumn{3}{|c|}{\textbf{en-US-Wavenet-E}} & \multicolumn{3}{|c|}{\textbf{en-US-Wavenet-I}} \\ \cline{2-10}
        & \textbf{Open} & \textbf{Wall} & \textbf{Small} & \textbf{Open} & \textbf{Wall} & \textbf{Small} & \textbf{Open} & \textbf{Wall} & \textbf{Small} \\
        \hline
        \textbf{Wakeword} & \zz{10} & \zz{10} & \zz{10} & \zz{10} & \zz{10} & \zz{10} & \zz{8} & \zz{8} & \zz{10} \\
        \textbf{``Hello''} & \zz{10} & \zz{10} & \zz{10} & \zz{10} & \zz{10} & \zz{10} & \zz{2} & \zz{6} & \zz{9} \\
        \textbf{``What time is it?''} & \zz{10} & \zz{10} & \zz{10} & \zz{10} & \zz{6} & \zz{6} & \zz{4} & \zz{7} & \zz{10} \\
        \textbf{``Turn off the light''} & \zz{4} & \zz{8} & \zz{9} & \zz{6} & \zz{8} & \zz{10} & \zz{2} & \zz{6} & \zz{10} \\
        \textbf{``Open Mask Attack''} & \zz{0} & \zz{4} & \zz{6} & \zz{0} & \zz{1} & \zz{0} & \zz{0} & \zz{6} & \zz{2} \\
        \textbf{``Call mom''} & \zz{2} & \zz{8} & \zz{8} & \zz{0} & \zz{4} & \zz{6} & \zz{1}& \zz{6} & \zz{8} \\
        \textbf{``Call 1234567890''} & \zz{0} & \zz{0} & \zz{0} & \zz{0} & \zz{0} & \zz{0} & \zz{0} & \zz{0} & \zz{0} \\
        \hline
    \end{tabular}
    }
\end{table}

\begin{table}[hbt!]
    \caption{Enhancement of TTS Commands Exploiting FVV}
    \label{bugabusepvc}
    \centering
    \resizebox{\columnwidth}{!}{%
    \begin{tabular}{ |c|c c c|c c c|c c c| }
        \hline
        \multirow{3}{*}{\textbf{TTS Voice Command}} & \multicolumn{9}{|c|}{\textbf{Google TTS Voice Profile and Scenario}} \\ \cline{2-10}
        & \multicolumn{3}{|c|}{\textbf{en-US-Wavenet-A}} & \multicolumn{3}{|c|}{\textbf{en-US-Wavenet-E}} & \multicolumn{3}{|c|}{\textbf{en-US-Wavenet-I}} \\ \cline{2-10}
        & \textbf{Open} & \textbf{Wall} & \textbf{Small} & \textbf{Open} & \textbf{Wall} & \textbf{Small} & \textbf{Open} & \textbf{Wall} & \textbf{Small} \\
        \hline
        \textbf{``Open Mask Attack''} & \zz{8} & \zz{4} & \zz{10} & \zz{0} & \zz{6} & \zz{4} & \zz{6} & \zz{6} & \zz{10} \\
        \textbf{``Call mom''} & \zz{10} & \zz{8} & \zz{9} & \zz{0} & \zz{4} & \zz{6} & \zz{8} & \zz{8} & \zz{9} \\
        \textbf{``Call 1234567890''} & \zz{6} & \zz{4} & \zz{10} & \zz{0} & \zz{2} & \zz{2} & \zz{0} & \zz{1} & \zz{0} \\
        \hline
    \end{tabular}
    }
\end{table}

Table~\ref{pvctable} reports the results of the best performing voice profiles. The table indicates the reliability of selected commands, with a score ranging from 0 to 10 according to the success rate of the command. For example, a command with 23\% success rate would score 3, while a command with 97\% success rate would score 10. Commands that never succeeded score 0. Each command was played 12 times on average, for each scenario. It can be observed that commands generated with the voice profile A perform better than the others in almost every scenario, and that the ``small'' scenario is generally the one with higher command reliability, due to the reflection of the soundwaves caused by multiple nearby obstacles. We also find that only profiles ``en-US-Wavenet-A'' and ``en-US-Wavenet-I'' can reliably open Mask Attack, while none of them can successfully dial a phone number.\footnote{Note that ``1234567890'' is a placeholder: a real phone number was used in the tests.} We verified that this is due to the volume being turned down after the wakeword recognition, which does not enable Alexa to properly recognize longer commands. Nonetheless, since the wakeword recognition happens with a short delay, voice profiles that speak fast are able to pronounce more words before the volume turn down, hence they can issue longer commands more reliably than other profiles.

\begin{figure}[!hbt]
    \centering
    \includegraphics[scale=0.6]{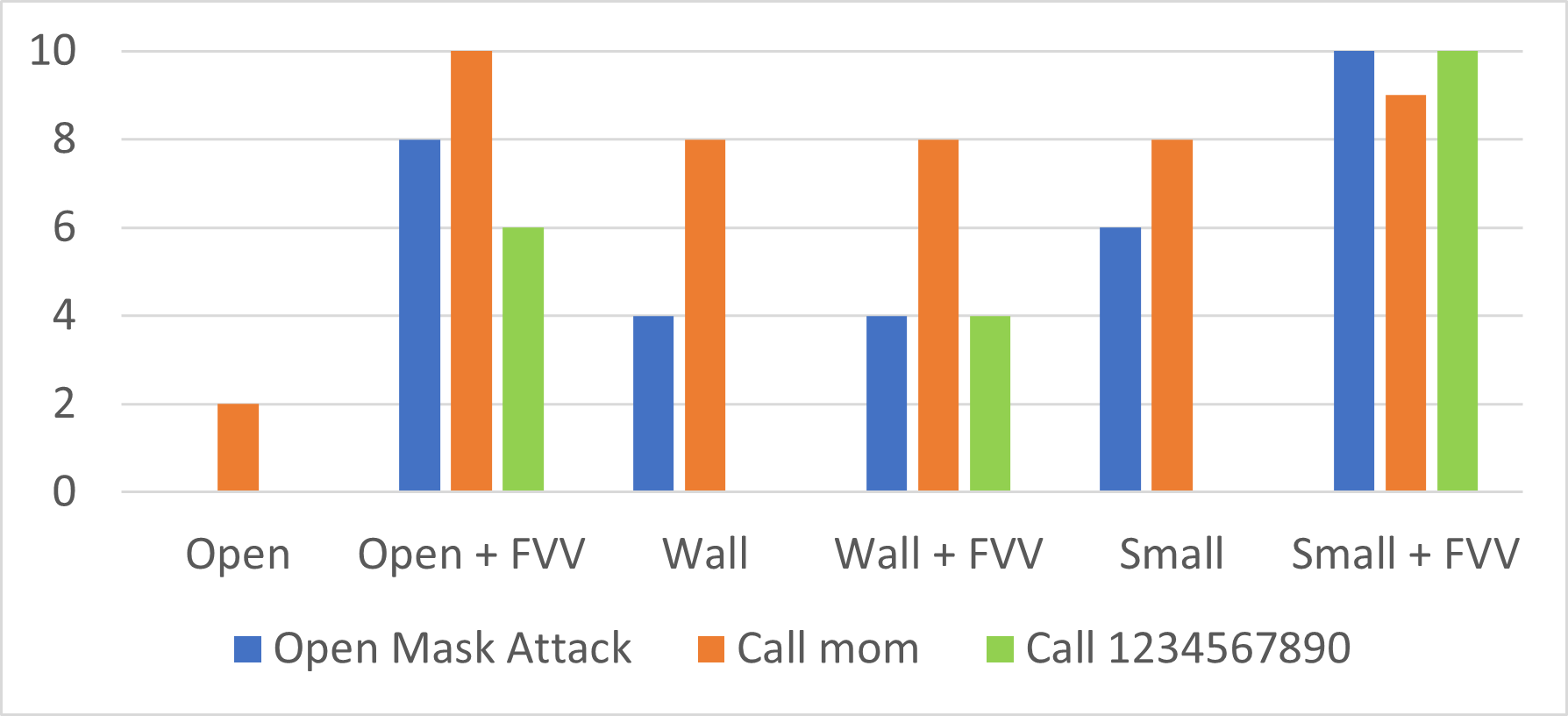}
    \caption{Comparison Between Normal Self-Issue and Enhanced Self-Issue With FVV (en-US-Wavenet-A)}
    \label{fve-enhance}
\end{figure}

\mypar{Full Volume Vulnerability}
Recall that in Section~\ref{exploitation} we introduced the Full Volume Vulnerability, which can be used to increase the success rate of self-issued commands, due to the fact that, after using it, the volume is not turned down after the subsequent wakeword recognitions. In Table~\ref{bugabusepvc} we re-evaluate the reliability of the commands that performed poorly in Table~\ref{pvctable}, issuing them after the FVV. Each command was played 12 times on average, for each scenario. Comparing these results with those in Table~\ref{pvctable}, we observe that performance for some commands has been dramatically enhanced, as the number dial with profile A in the open and small space scenario. Figure~\ref{fve-enhance} shows a direct comparison between success rates exhibited with and without the FVV, and we can see that performance of commands issued with FVV is always equal or higher than the normal ones'. Recall that the attacker can choose the samples to use during the attack, hence they can select the best ones and loop them during the \attackname{} attack.

\mypar{Effectiveness Over Time}
We observed that commands generated with certain voice profiles (e.g., ``en-US-Wavenet-I'') lose effectiveness over time: in fact, if they are used too many times in a short timespan, the Echo device stops recognizing them. We believe this is a defense against replay attacks. To this end, during our tests, \attackname{} slightly edited the voice pitch after each attempt to be able to re-issue commands. In addition to this, we verified that the commands regain their effectiveness if the device has been moved or after few minutes of inactivity: this last solution could also be adopted by an attacker as they would not need to issue a high-rate of commands. Additionally, the adversary can cycle through the commands they have generated, instead of re-using the same sample every time.

\mypar{Effectiveness of Self-Triggering with Low Volume Level}
We tested TTS commands while no skill was running in the background, for different volume levels and in different scenarios. For this test as well, each command was played 12 times on average, for each scenario -- results are shown in Table~\ref{volumetable}. We observe that the success rate of the commands is not proportional to their volume (higher volume does not mean higher success rate). During the tests, we noted a degradation in efficacy of the attack if the volume falls under 3. In these cases, we observed that the further volume reduction caused by the wakeword recognition, or by a skill running on the Echo device, rendered the audio completely inaudible, hence, it was not possible to reliably self-issue long commands anymore for most voice profiles.

\mypar{Effectiveness in Presence of Another Audio Stream}
We tested a scenario where the legitimate user tries to start another audio stream while the Echo device is connected to an attack vector, for example by saying ``\textit{Alexa, play Despacito on Spotify}'', or by connecting a device via Bluetooth. Table~\ref{musicprioritytable} illustrates all possible situations: we observed there are three outcomes, namely, the attack vector is disconnected permanently (stop: \faStop), the attack vector is temporarily disconnected and is reconnected after the user has finished listening to their own music (pause: \faPause), or the attack vector keeps the connection, and the user is not allowed to play their music (play: \faPlay). The favourable scenarios for \attackname{} are those marked with the ``play'' and ``pause'' symbols, since the attack vector is not disconnected, while those marked with the ``stop'' symbol are not favourable as they disconnect Echo from the attack vector.

\begin{table}[h!]
    \caption{Effectiveness of Self-Triggering (``\textit{Echo, what time is it?}'') When No Skill is Running in the Background}
    \label{volumetable}
    \begin{center}
    \resizebox{\columnwidth}{!}{%
    \begin{tabular}{ |c|c|c c c|c c c|c c c|c c c|c c c|c c c|}
        \hline
        \multicolumn{2}{|c|}{\textbf{Voice Profile}} & \multicolumn{18}{|c|}{\textbf{Volume Level and Scenario}} \\
        \hline
        \multirow{2}{*}{\textbf{Name}} & \multirow{2}{*}{\textbf{Gen.}} & \multicolumn{3}{|c|}{\textbf{6+}} & \multicolumn{3}{|c|}{\textbf{5}} & \multicolumn{3}{|c|}{\textbf{4}} & \multicolumn{3}{|c|}{\textbf{3}} & \multicolumn{3}{|c|}{\textbf{2$^*$}} & \multicolumn{3}{|c|}{\textbf{1$^*$}} \\ \cline{3-20}
        & & \textbf{O} & \textbf{W} & \textbf{S} & \textbf{O} & \textbf{W} & \textbf{S} & \textbf{O} & \textbf{W} & \textbf{S} & \textbf{O} & \textbf{W} & \textbf{S} & \textbf{O} & \textbf{W} & \textbf{S} & \textbf{O} & \textbf{W} & \textbf{S} \\
        \hline
        \textbf{A} & \faMale & \zz{10} & \zz{10} & \zz{10} & \zz{10} & \zz{10} & \zz{10} & \zz{10} & \zz{10} & \zz{10} & \zz{9} & \zz{10} & \zz{10} & \zz{2} & \zz{10} & \zz{10} & \zz{1} & \zz{10} & \zz{10} \\
        \textbf{B} & \faMale & \zz{0} & \zz{0} & \zz{6} & \zz{0} & \zz{6} & \zz{6} & \zz{0} & \zz{10} & \zz{6} & \zz{0} & \zz{10} & \zz{6} & \zz{0} & \zz{0} & \zz{0} & \zz{0} & \zz{0} & \zz{0} \\
        \textbf{C} & \faFemale & \zz{0} & \zz{4} & \zz{4} & \zz{2} & \zz{4} & \zz{6} & \zz{0} & \zz{6} & \zz{6} & \zz{0} & \zz{8} & \zz{6} & \zz{2} & \zz{0} & \zz{6} & \zz{1} & \zz{0} & \zz{5} \\
        \textbf{D} & \faMale & \zz{0} & \zz{2} & \zz{6} & \zz{1} & \zz{2} & \zz{8} & \zz{2} & \zz{8} & \zz{10} & \zz{0} & \zz{4} & \zz{10} & \zz{0} & \zz{0} & \zz{10} & \zz{0} & \zz{0} & \zz{10} \\
        \textbf{E} & \faFemale & \zz{0} & \zz{2} & \zz{6} & \zz{10} & \zz{6} & \zz{6} & \zz{10} & \zz{9} & \zz{10} & \zz{8} & \zz{10} & \zz{10} & \zz{10} & \zz{0} & \zz{10} & \zz{10} & \zz{0} & \zz{0} \\
        \textbf{F} & \faFemale & \zz{0} & \zz{2} & \zz{10} & \zz{0} & \zz{2} & \zz{10} & \zz{0} & \zz{10} & \zz{10} & \zz{0} & \zz{10} & \zz{10} & \zz{0} & \zz{0} & \zz{10} & \zz{0} & \zz{9} & \zz{10} \\
        \textbf{G} & \faFemale & \zz{0} & \zz{2} & \zz{2} & \zz{2} & \zz{2} & \zz{10} & \zz{0} & \zz{2} & \zz{10} & \zz{2} & \zz{10} & \zz{10} & \zz{0} & \zz{2} & \zz{6} & \zz{1} & \zz{0} & \zz{10} \\
        \textbf{H} & \faFemale & \zz{0} & \zz{0} & \zz{1} & \zz{2} & \zz{4} & \zz{6} & \zz{10} & \zz{2} & \zz{10} & \zz{0} & \zz{2} & \zz{10} & \zz{0} & \zz{10} & \zz{10} & \zz{0} & \zz{0} & \zz{10} \\
        \textbf{I} & \faMale & \zz{4} & \zz{5} & \zz{10} & \zz{4} & \zz{7} & \zz{10} & \zz{0} & \zz{10} & \zz{10} & \zz{0} & \zz{10} & \zz{10} & \zz{0} & \zz{10} & \zz{10} & \zz{0} & \zz{10} & \zz{10} \\
        \textbf{J} & \faMale & \zz{0} & \zz{6} & \zz{6} & \zz{0} & \zz{4} & \zz{6} & \zz{0} & \zz{2} & \zz{10} & \zz{0} & \zz{0} & \zz{10} & \zz{0} & \zz{0} & \zz{10} & \zz{0} & \zz{0} & \zz{10} \\
        \hline
    \end{tabular}
    }
    \end{center}
    \footnotesize{\textbf{Name:} The full name of the voice profile is ``en-US-Wavenet-X'', where X is the letter shown in the column.   \textbf{*}When the volume is set at 1 or 2, the further volume turn down caused by Echo recognizing the wakeword makes the played audio inaudible. Some commands succeed nonetheless due to the fact that the words ``what time'' manage to be played before the volume is muted, allowing the command to be interpreted correctly. However, not all commands can be shortened enough to make them work with volume 1 and 2.}
\end{table}    

\begin{table}[h!]
    \caption{Behaviour of Echo When the Adversary is Already Streaming Commands and the User Asks Echo to Play Music}
    \label{musicprioritytable}
    \begin{center}
        \begin{tabular}{ |c|c|c|c| }
        \hline
        \multicolumn{2}{|c|}{\textbf{Pre-conditions}} & \multicolumn{2}{|c|}{\textbf{User attempts to...}} \\
        \hline
        \textbf{Atk Vector} & \textbf{VMA on?} & \textbf{\textit{``Play Music''}} & \textbf{Connect BT} \\
        \hline
        \textbf{Radio} & \rednocheck & \faStop & \faPause \\
        \textbf{Radio} & \greenyescheck  & \faPlay & \faPause \\
        \textbf{Bluetooth} & \rednocheck & \faPlay* & \faStop \\
        \textbf{Bluetooth} & \greenyescheck  & \faPlay & \faStop \\
        \hline
    \end{tabular}
    \end{center}
    \footnotesize{$^*$In this case, both audio tracks are played at once. However, if the attacker stops their track and plays it again, they gain priority over the radio station, which gets muted.}
\end{table}

\section{Feasibility and Impact Assessment}\label{sec:Feasibility}
Here, we first assess the feasibility of using \attackname{} in real scenarios, by analysing the results of a field study on three different voluntary households (\S\ref{userstudy}). Secondly, we detail a set of attacks that can be performed on the victim's Echo device via \attackname{}, also showing their impact on the user's physical safety and privacy (\S\ref{impactsection}).

\subsection{Field Study on Echo-Equipped Households}
\label{userstudy}

To assess the feasibility and impact of \attackname{} in real scenarios, we performed a field study with three different households. All participants in the three households possessed an Amazon Echo device at various level of expertise. For practical and ethical reasons, during these experiments, we asked the participants to use our own device, and we placed this device in the same position and with the same orientation of the participants' device. All the users in the three households were informed about the replacement of the device and they were told we would perform some usability tests by observing their interactions with our device. Before the tests, we did not specify the users the exact goals of the tests as not to introduce bias. For ethics of this study, see Appendix~\ref{responsibledisclosure}.

\mypar{Stealth Attack Vector Connection} \label{teststealth} The goal of this test is to check whether is it possible, for an attacker who temporarily has access to the household, to connect to Echo via Bluetooth without being detected. In a real scenario this could be a professional who has to repair something in the house, or someone who was invited over by the legitimate user. In all three households, we were able to connect to Echo without anyone realizing it: this can be trivially done if no one is in the room for a very short time frame. We set the volume of the Echo device to 1 (\textit{``Echo, volume 1''}), then we turned on the Bluetooth (\textit{``Echo, turn on Bluetooth''}) and we connected to it. We interrupted all Alexa's status messages by pressing the action button repeatedly. The whole process lasted 25 seconds on average.

\mypar{Command Self-Issue Perception} The goal of this test is to understand if people in the household are always able to hear the TTS commands, or if there are some conditions, such as their position in the house and the device's volume, that allow the attacker to self-issue commands even when the user is at home and awake. We set the device to the volume used by the different households (respectively 3, 5 and 6) and we told the users to go in different parts of the house.
We then self-issued a command to the Echo device, let Echo play the normal response, and then we asked the users if they did hear something. Regardless of the volume, all users in the same room and in adjacent rooms did hear both the command and the reply correctly. Only in the household with volume 3, in an adjacent room with the door closed, the user reported they thought it was someone talking outside of the window, and they could not infer what was said.
Finally, no user in any household was able to hear neither the command, nor the reply, if they were in a non-adjacent room, although a user in the household with volume 6 reported they thought Alexa had said something.

\mypar{User Behaviour After Bluetooth Connection} When the adversary reconnects to an Echo device they have already paired their device with, they will skip the pairing process. However, the connection prompts Echo to generate a ``successful connection'' audio message that the user can hear -- hence, the goal of this test is to assess what would the users do when hearing the message. We connected our device to Echo surreptitiously, when the users were near the device.
When asked separately if they could tell what happened, half of them could infer that we had connected a device to Echo via Bluetooth. However, they did not perceive it as a malicious action. Only one user tried to understand what happened by asking a series of questions to Alexa, for example ``what happened?'' and ``are you connected to something?'', however, the device did not reveal any ongoing Bluetooth connection.

\mypar{Command Self-Issue from Outside} The goal of this test is to check whether is it possible to issue commands from outside the houses, after a device has been paired with Echo. We disconnected our paired device from Echo and we tried to reconnect once outside. We successfully managed to reconnect our device in all three cases, also succeeding in self-issuing voice commands. In two cases, the distance from the device was 8m, with two walls between us and the device. In another household, the distance was 3m, with only one wall separating us from the device.
    
\mypar{Interaction with the Mask Attack Skill} The goal of this test is to assess the behaviour of people interacting with the Mask Attack skill. In particular, we wanted to understand whether they were able to infer that an attack was in place. We told the users we wanted to observe their interactions with the Echo device, and that they had total control over it -- they could change the volume, mute the microphone, issue any command, etc. They were then presented the device with the Mask Attack skill already opened, unknowingly to them.
We did not use the ``Echo, go on'' command to refresh the timeout timer during this test.
All users noticed that Echo was slower in replying to the commands. When issuing commands with very long replies, the Oracle could not reply within the allocated five seconds, hence Mask Attack sent back the reply to the last issued command. However, when they issued the command again, they could hear the correct reply since the Oracle had the time to retrieve the answer, convert it to text and store it into the database. All of them reported that they attributed this behaviour to a bug. One of the users noticed the blue light blinking upon receiving an incorrect reply, due to an erroneous transcription of a date by the Speech-To-Text service. The user then restarted the device, reporting that they often get weird replies from Alexa (a feeling shared by the other participants as well), however, they thought it was safer to restart the device because of the blinking light. All the other users participating in the study did not notice the blinking light, or perceived it as normal. Additionally, all the other users kept interacting with the device until we stopped the experiment.

\mypar{Summary of the Field Study's Results} Table~\ref{table:userstudy} summarizes the results of the field study. As we can see, it is possible to quickly and stealthily set up a Bluetooth connection with Echo, and we also verified that it was possible to issue commands from outside the household in all the cases. This confirms the dangerousness of the attack vector. As expected, the study confirms the limitation of the audible payloads, hence, the adversary needs to carefully plan when to self-issue commands to avoid alerting the users.
Nonetheless, all other tests show that \attackname{} can successfully start and keep running, as most of the users did not perceive anomalies as malicious.

\begin{table}[hbt!]
    \caption{Field Study Results}
    \label{table:userstudy}
    \centering
    \footnotesize
    \resizebox{\columnwidth}{!}{
    \begin{tabular}{ l l l }
        
        \Xhline{2\arrayrulewidth}
        \textbf{Test} & \textbf{Observation Item} & \textbf{Result} \\
        \hline
        
        \multirow{2}{*}{Stealth BT} & Time Needed & \textcolor{darkgreen}{25s on average} \\
        & Users Aware & \textcolor{darkgreen}{0\% of users} \\
        \hline
         
        \multirow{4}{*}{SI Perception} & Same Room & \textcolor{red}{100\% of users could hear} \\
        & 3.5m and Wall & \textcolor{red}{100\% of users could hear} \\
        & 3.5m, Wall and Door & \textcolor{orange}{66\% of users could hear} \\
        & Non Adj. Room (7m+) & \textcolor{darkgreen}{0\% of users could hear} \\
        \hline
        
        \multirow{3}{*}{BT Connect} & Perceived as Malicious & \textcolor{darkgreen}{0\% of users} \\
        & Inferred What Happened & \textcolor{orange}{50\% of users}\\
        & Took Some Actions & \textcolor{darkgreen}{16\% of users}\\
        \hline
        
        \multirow{2}{*}{Outdoor Self-Issue} & Successful Pairing & \textcolor{darkgreen}{100\% of households}\\
        & Successful SI & \textcolor{darkgreen}{100\% of households} \\
        \hline
        
        \multirow{6}{*}{Mask Attack} & Noticed Slow Behaviour & \textcolor{red}{100\% of users} \\
        & Noticed Wrong Replies & \textcolor{red}{100\% of users} \\
        & Noticed Blinking Light Ring & \textcolor{darkgreen}{16\% of users} \\
        & Perceived as Malicious & \textcolor{darkgreen}{16\% of users} \\
        & Switched Off the Device & \textcolor{darkgreen}{16\% of users} \\
        & Intercepted Commands & \textcolor{darkgreen}{100\% (41/41)} \\
        \Xhline{2\arrayrulewidth}

    \end{tabular}
    }
\end{table}

\subsection{\attackname{} Impact Assessment}
\label{impactsection}
We now detail the types of malicious actions \attackname{} allows an attacker to perform, and we assess their success rate.

\mypar{Control Other Smart Appliances} \attackname{} allows an attacker to control other smart appliances that are connected to Echo. This action can undermine physical safety of the user, for example, when turning off the lights during the evening or at night time, turning on a smart microwave oven, setting the heating at a very high temperature or even unlocking the smart lock for the front door. We were able to turn off the lights in one of the authors' house 93\% of the times using the FVV. In these scenarios, sometimes Echo might repeat the name of the device the attacker wants to turn off as a form of double-check. In this case, the adversary only has to always append a ``yes'' approximately six seconds after the request to be sure that the command will be successful.

\mypar{Call Any Phone Number of Attacker's Choice} The attacker can make the victim Echo device call a phone number controlled by them, effectively eavesdropping on what is being said in proximity of the Echo device. While it is true that the user could see that the light on top of the device turns green (or on the bottom of the device in 4th Generation devices), it may take a while for the user to notice it, depending on the position of the device. Additionally, unexperienced users could not know what the green light means, hence ignoring it. In fact, among users who answered the survey, only 27\% knew that it is related to ongoing calls. During the test, we exploited \attackname{} to call two phone numbers owned by the authors of this work, by placing the device in different locations, and we succeeded 73\% of the times when using the FVV. In the remaining 27\%, some numbers were incorrectly interpreted by the Echo device.

\mypar{Buying Unwanted Items on Amazon Using the Victim's Account} Although the user would be notified of the purchase via email, and be able cancel the malicious order, this would be detrimental for the victim's user experience on Amazon, and they could lose trust in the company. The attacker can also delete items that the user had previously put in the shopping cart. We were able to perform both actions using the command self-issue, and we achieved 100\% success rate when using the FVV. Echo did not correctly understand what product we wanted to buy once (20\%), however, another article was placed in the shopping cart nonetheless, and we were able to buy it.

\mypar{Tampering the User's Linked Calendar} If the user had previously linked their online calendar, the attacker can exploit \attackname{} to add, move or delete events from the user's calendar. We were able to perform all these actions, via self-issue, on a Google Calendar linked to Alexa. With FVV, we succeeded 88\% of the times.

\mypar{Impersonate Other Skills and the VPA} The attacker can exploit \attackname{} to start any skill of their choice, including Mask Attack to perform our instance of the VMA. Impact of VMA attacks is already thoroughly discussed in other works~\cite{dangerousskills, lyexa}, and they can seriously undermine the victim user's privacy. In fact, they could lead the user to disclose their passwords or security PINs, personal data (e.g., home address, name and surname), or even sensitive data, such as health status, religious belief and sexual preferences. During our tests with Echo users, no one noticed that the Mask Attack skill was running, although all users could understand that Echo was experiencing some problems due to the delay in the replies. In the tests, only one user noticed the blinking light, and thought it was safer to shutdown the Echo device. All the other users kept interacting normally with Echo.

\mypar{Retrieve User Utterances} As Mask Attack stores all intercepted commands in a database, this would allow the adversary to extract private information, gather information on used skills and infer user habits, e.g., to calculate when it is safer to self-issue commands. During the field study, we were able to retrieve all utterances made by the users -- as explained in \S\ref{userstudy}.

\mypar{Summary of the Impact Assessment} Table~\ref{table:impact} summarizes our findings, where green indicates an optimal success rate (we set the threshold at $\geq$80\%) and yellow indicates a medium success rate.\footnote{As explained in Appendix~\ref{responsibledisclosure}, during all the tests we have only used accounts of the authors of this work.}

\begin{table}[hbt!]
    \caption{Impact Assessment Summary}
    \label{table:impact}
    \scriptsize
    \begin{center}
    \resizebox{\columnwidth}{!}{
    \begin{tabular}{ l l r l l }
        
        \Xhline{2\arrayrulewidth}
        \textbf{Threat} & \multicolumn{2}{c}{\textbf{Success Rate}} & \textbf{Confirm?}$\dagger$ & \textbf{Notes} \\
        \hline
        
        Control Smart Dev. &  \textcolor{darkgreen}{14/15} & *\textcolor{darkgreen}{(93.3\%)} & Sometimes & Tested with lights only \\
        
        Call Phone Number & \textcolor{orange}{11/15} & *\textcolor{orange}{(73.3\%)} & Yes & 14 digits including prefix \\
        
        Buy Items & \textcolor{darkgreen}{5/5} & *\textcolor{darkgreen}{(100\%)} & Yes & 80\% (4/5) items recognized \\
        
        Tamper Calendar & \textcolor{darkgreen}{8/9} & *\textcolor{darkgreen}{(88.8\%)} & Sometimes & Did not ask when moving event \\
        
        VMA & - & - & - & See Table~\ref{table:userstudy} \\
        
        Retrieve Utterances & \textcolor{darkgreen}{41/41} & \textcolor{darkgreen}{(100\%)} & - & See Table~\ref{table:userstudy} \\

        \Xhline{2\arrayrulewidth}

    \end{tabular}
    }
    \end{center}
    \textbf{*} Indicates the success rate was achieved using the FVV. | $\dagger$ indicates if the command requires the attacker to self-issue a ``Yes'' after a few seconds to answer the confirmation request.
\end{table}

With these tests, we demonstrated that \attackname{} can be used to give arbitrary commands of any type and length, with optimal results -- in particular, an attacker can control smart lights with a 93\% success rate, successfully buy unwanted items on Amazon 100\% of the times, and tamper a linked calendar with 88\% success rate. Complex commands that have to be recognized correctly in their entirety to succeed, such as calling a phone number, have an almost optimal success rate, in this case 73\%. Additionally, results shown in Table~\ref{table:userstudy} demonstrate the attacker can successfully setup a Voice Masquerading Attack, via our Mask Attack skill, without being detected and all issued utterances can be retrieved and stored in the attacker's database, namely 41 in our case.

\section{Limitations Assessment}
Here, we first discuss the limitations of \attackname{} (\S\ref{limitations}), and then we report the results of a survey submitted to a study group of 18 Amazon Echo users, to evaluate the limitations' effectiveness (\S\ref{surveyresults}). 

\subsection{Limitations}\label{limitations}

While \attackname{} can issue arbitrary commands for a prolonged time on Echo, some events can terminate the attack. We list and compare them with the results of the field study (\S\ref{userstudy}) and of a user survey (\S\ref{surveyresults}), showing that the likelihood for them to take place is minimal.

\mypar{The Echo Device is Unplugged from Power} Radio stations, and in general played audio tracks, are not automatically played again when the Echo device is restarted, and this would disconnect Echo from the attack vector. However, only 27\% of the users of the survey ever restarted their Echo device, and only 6\% do it systematically.
    
\mypar{The User Says ``Alexa, Stop''} Such command would close the radio station, disconnecting Echo from the attack vector. However, if the Mask Attack skill is running, the command ``Alexa, stop'' would close the skill instead, so the user would have to say the command two times in a row to terminate the radio station. This does not apply when using the Bluetooth attack vector, as this command does not stop the stream. We argue that, realistically, the user does not say this command without a reason. During our experiments, no users tried to issue this command.
    
\mypar{Headphones Connected to Echo} \attackname{} would not succeed since the malicious commands would be played through the headphones. In our study group, no user ever tried to connect them to Echo.
    
\mypar{Echo's Microphone is Turned Off} Echo Dot includes a button to turn off its microphone. If pressed, payloads cannot exploit the self-issue vulnerability since the microphone would not capture anything. Nearly 89\% of the users never turn off Echo's microphone, or they do it rarely. The remaining 11\% claimed to do it sometimes.

\mypar{Payloads are Audible} Although adversarial commands can be used with \attackname{}, they are not yet very reliable when self-issued (see Appendix~\ref{attackwithhidden} for details). To deploy a real attack, the adversary has to use TTS commands, that can be heard by nearby users up to 4.5 metres on average, as shown in Section~\ref{userstudy}. Because the attacker does not know if the legitimate user is nearby, they might want to issue arbitrary commands during the night.
    
\mypar{Volume Buttons} While the Mask Attack skill is on, the only way for the user to change the volume is by using the physical buttons on the Echo device, since the skill is intercepting voice commands: this could make the user suspicious. This would not affect 27\% of the users in our study group, who claimed to use manual commands to change Echo's volume.
    
\mypar{Echo's Light Ring Turns Green During a Call} When the attacker makes Echo call a phone number they control to eavesdrop on the household, the user could notice the light ring become green. Depending on the position of the device, this could take some time. Additionally, the survey reports that only 27\% of the users know what the green light means.
    
\mypar{Echo's Light Ring Blinks when Reading} While Alexa is talking, Echo's light ring slowly blinks. When the Mask Attack skill is open, the light ring keeps on blinking as Alexa is reading the \texttt{break} tags issued by the skill. As for the green light, depending on the position of the device, the users might not notice it. In fact, during our field study, only one user noticed the blinking light.
    
\mypar{Inaccuracy of Reply Retrieval} Some skills use pre-recorded audio tracks, such as the voice of a journalist. Mask Attack will read these tracks with Alexa's voice. Additionally, \textit{by default}, skills cannot access specific data on the user account, e.g. shopping lists, unless the user grants them the appropriate permissions via the Alexa companion app. For such commands, Mask Attack will send a feasible reply anyway, however, it will not be the \textit{correct} reply.
    
\mypar{Delay Before the Reply} The infrastructure behind Mask Attack needs some time to convert the user utterance into audio and, vice-versa, the reply into text. All users noticed this delay, however, none of them perceived it as malicious. 

\subsection{Evaluation of the Limitations}\label{surveyresults}

We next evaluate how realistically the limitations of \attackname{} discussed in \S\ref{limitations} would reduce the effectiveness of \attackname{}. To this end, we created a detailed survey composed of twelve questions, clustered in three categories: usage information, scenario recognition and limitations assessment. The first part of the survey was aimed at collecting information on how people use their Echo device, such as frequency of use and commonly issued commands. The second part of the survey was aimed at understanding the most common placement of the device and the average volume of the device. The third part of the survey directly assesses how limitations of \attackname{} are actually used in practice common among Echo users. We submitted the survey to a study group of 18 Echo owners, with different levels of ability in using their Echo device. Some users owned more than one device, and they were asked to answer the survey choosing one of their devices randomly.

Table~\ref{table:limitations} summarizes the evaluation of the limitations, based on the results of the survey\footnote{Questions in the survey and results of the answers can be found in Appendix~\ref{app:survey}.} and of the field study. We can see that all limitations that might stop the attack have a very low likelihood, hence they are not a real threat for \attackname{}. Among events that might alert users, results show that it is not very likely that users realize that a phone call is taking place, so an adversary would probably be able to eavesdrop on the users. However, the fact that users can hear the self-issued payloads is confirmed to be the main limitation of the attack, hence the adversary needs to carefully choose when to send commands to Echo. All other events in the table are mostly ignored by Echo users, and only expert ones would be possibly alerted by their presence. Hence, we argue that most limitations for the \attackname{} attack remain theoretical.

\begin{table}[hbt!]
    \caption{Evaluation of the Limitations}
    \label{table:limitations}
    \footnotesize
    \begin{centering}
    \begin{tabular}{ l l r l }
        
        \Xhline{2\arrayrulewidth}
        \textbf{Limitation} & \multicolumn{2}{c}{\textbf{Likelihood}} & \textbf{Impact} \\
        \Xhline{2\arrayrulewidth}
        
        Echo's microphone is muted & \textcolor{darkgreen}{Low} & (11\%) & \textcolor{red}{$\bullet\bullet\bullet\ \bullet$} \\
        
        Echo is unplugged from power & \textcolor{darkgreen}{Low} & (6\%) & \textcolor{red}{$\bullet\bullet\bullet\ \bullet$}  \\

        User says ``Alexa, stop'' & \textcolor{darkgreen}{Low} & (0\%) & \textcolor{red}{$\bullet\bullet\bullet\ \bullet$}  \\
        
        Headphones are connected to Echo &  \textcolor{darkgreen}{Low} & (0\%) & \textcolor{red}{$\bullet\bullet\bullet\ \bullet$}  \\
        
        User hears payload (same room) & \textcolor{red}{High} & (100\%) & \textcolor{darkorange}{$\bullet\bullet\bullet$} \\
        
        User hears payload (adj. room) & \textcolor{orange}{Medium} & (66\%) & \textcolor{darkorange}{$\bullet\bullet\bullet$} \\
        
        User notices ongoing call (green light) & \textcolor{orange}{Medium} & (27\%) & \textcolor{darkorange}{$\bullet\bullet\bullet$} \\
        
        User hears payload (non-adj. room) & \textcolor{darkgreen}{Low} & (0\%) & \textcolor{darkorange}{$\bullet\bullet\bullet$} \\
        
        User notices inaccuracies in replies & \textcolor{red}{High} & (100\%) & \textcolor{orange}{$\bullet\ \bullet$} \\
        
        User cannot change volume by voice & \textcolor{orange}{Medium} & (73\%) & \textcolor{orange}{$\bullet\ \bullet$} \\
        
        User realizes light ring is blinking & \textcolor{darkgreen}{Low} & (16\%) & \textcolor{orange}{$\bullet\ \bullet$} \\

        User notices delay before the reply & \textcolor{red}{High} & (100\%) & \textcolor{darkgreen}{$\bullet$} \\
 
        \Xhline{2\arrayrulewidth}

    \end{tabular}
    
    \textcolor{red}{$\bullet\bullet\bullet\ \bullet$}: It might stop the attack. | \textcolor{darkorange}{$\bullet\bullet\bullet$}: It may alert any user. | \textcolor{orange}{$\bullet\ \bullet$}: It is ignored by most users, but can still alert an expert user | \textcolor{darkgreen}{$\bullet$}: It is ignored by all users.
    \end{centering}
\end{table}

\section{Countermeasures}\label{countermeasures}

We now illustrate some countermeasures that can be applied to the VPA to mitigate \attackname{}'s threats regardless of the used payload.

\mypar{Self-Generated Wakeword Suppression} Similarly to the process of web and database input sanitization, VPAs should ignore commands coming from their own speakers. This could require the implementation of a wakeword detection system for the audio outputted by Echo: if a wakeword is detected within the played audio, such wakeword is not considered valid~\cite{patent1, patent3}.
Alternatively, directional audio signals might be analyzed, to check whether the wakeword comes from a single or multiple directions, the latter indicating that the wakeword is being self-issued~\cite{patent2}. It is possible that Echo Dot already implements a similar technology, as an array consisting of four microphones can be found within the device~\cite{echoteardown}, however, we have not been able to confirm this.

\mypar{Liveness Detection} Detecting whether the source of the captured audio is a human or an electronic speaker would defeat \attackname{}. For instance, it could be possible for the VPA to analyze the spectrum of the captured audio to identify low frequency signals that cannot be emitted by the human voice, but that are distinctive of an electronic speaker \cite{helloisitme}. Alternatively, it is possible to distinguish the airflow generated by a real user from the one generated from an electronic speaker~\cite{secureyourvoice}. However, this approach requires an additional airflow sensor, which Echo Dot currently does not have.

\mypar{Automatic Speaker Verification} 
Currently, Alexa gives users the possibility to perform a voice training, however, this mechanism is not used as a security measure to lock actions because there exists a number of known attacks that might be able to trick Automatic Speaker Verification systems (ASV) into classifying a spoofed audio as legitimate~\cite{practicalhvc, kenansville, houdini, Yakura, kreuk}. The goal of an ASV is to verify that a given sentence has been pronounced by one of the users that the ASV system has learnt to recognize via a voice training process. Thanks to competitions such as ASVspoof~\cite{asvspoof19}, resilient ASV solutions are being developed (e.g., ASSERT~\cite{assert}), which could be used as a mitigation against \attackname{}. We argue that a more robust mitigation against \attackname{} would consist of an ASV that is trained with a corpus of audios recorded via a microphone array similar to the one used by Amazon Echo.\footnote{E.g., the Respeaker 4-Mic Array for Raspberry Pi, manufactured by the Seeed company.} This corpus should include (i) legitimate  commands recorded by different human speakers, (ii) synthetic commands generated via TTS, and (iii) replayed commands. In the last two cases, the electronic speaker playing these audio files should be placed directly under the microphone array to simulate the self-issue. In addition, the spoofed audios would have to be recorded by placing the microphone array in different locations, namely the open, wall and small scenarios described in \S\ref{threatmodel}. This corpus would allow the ASV solution to learn how to distinguish self-issued commands from the legitimate ones. 

\section{Related Work}
\label{relatedwork}

\mypar{Attacks Leveraging Self-Issue of Voice Commands} \citet{a11yattacks} leverage accessibility (a11y) tools made available on most OSes to identify two attacks that make use of voice commands self-issue. The first one works on Windows systems, and exploits the Windows Speech Recognition, which can be started by any process. By self-issuing some specific commands, any malware can escalate from low privileges to administrator, and can control the system via the voice channel. The second attack works on Android smartphones, and allows an attacker to bypass the Voice Authentication mechanism by recording and then replaying the authentication phrase, which is usually ``OK Google''. \citet{yourvoiceassistantismine} develop malware (VoicEmployer) that runs on Android and is able to self-issue voice commands using Google Voice Search when the user is not listening. Their attack (GVS-Attack) does not require any permission. \citet{monkeysays} use Google Firebase as a C\&C server and leverage sensors that require no permissions to infer if a smartphone is left unattended. If it is the case, the C\&C is contacted and the device issues TTS commands to itself and to nearby devices, if possible.

\mypar{Attacks Relying on Misinterpretation} \citet{skillsquatting} classify errors made by VPAs when interpreting a voice command into three categories: (i) \textit{homophones} are two words pronounced in the same way but with different spelling; (ii) \textit{compound words} can be split in their components, as in ``outdoors'' and ``out doors''; (iii) \textit{phonetic confusion} is the misclassification of one phoneme with a similar one, resulting in the transcription of a different word. The authors also introduce the concept of Skill Squatting Attack, an attack where Alexa opens a (potentially malicious) skill not meant by the user. \citet{alexaisthisskillsafe} analyze over 90,000 skills to find out that the Skill Squatting Attack is not being used systematically in the wild, and observe that multiple skills can have the same invocation name, hence, the user could activate a wrong skill.

\citet{dangerousskills} introduce the Voice Masquerading Attack (VMA), in which a malicious skill impersonates the VPA to deceive the user and exfiltrate personal data. During the VMA, the adversary could also pretend to invoke a skill if the user requests it, impersonating it and faking its termination when appropriate. \citet{missensebispham} analyze nonsense attacks on Google Assistant, showing that it is possible to trigger actions using words that are nonsensical to the human ear, but that are intepreted as valid commands by the VPA.

\mypar{Adversarial Attacks} \citet{sokfaultasr} analyze attacks and defenses on automatic speech and speaker recognition, and provide a taxonomy of threat models on the Voice Processing Systems. The authors also test the transferability property for adversarial commands, showing that it is still difficult to craft adversarial samples that are reliably transferable. \citet{carliniaudioadv} build targeted audio adversarial examples against Mozilla's DeepSpeech ASR implementation: their white-box attack is able to alter any waveform into another that is almost identical to the original, and gets classified by DeepSpeech as any other text chosen by the adversary. However, this attack is not feasible over-the-air. Devil's Whisper~\cite{devilwhisper} embeds hidden commands into songs, such as CommanderSong~\cite{commandersong} from which it stems. Devil's Whisper targets black-box systems using a transferability-based approach and two models to generate adversarial samples for most of the commercial ASR systems. \citet{lyexa} perform a man-in-the-middle attack (MITM), called Lyexa, which leverages a nearby malicious device to jam the VPA and alter user commands requested, e.g. by requesting the use of a malicious skill to perform any action. To this end, they leverage inaudible voice commands, as in DolphinAttack~\cite{dolphinattack}. An adversarial attack not leveraging audio files is discussed in the work by \citet{lightcommands}, which aims amplitude-modulated light towards an exposed microphone aperture in a VPA device and is able to inject arbitrary commands even when more than 100 metres away from the target device. Similarly, \citet{surfingattack} make use of a piezoelectric transducer to guide ultrasonic waves through a solid transmission media to inject inaudible commands.

\mypar{Comparison of \attackname{} with Similar Works}
\label{otherworks}
The self-issuing of voice commands on devices that use them as input has been already covered. In particular, two works \cite{a11yattacks, yourvoiceassistantismine} achieve similar results on Windows and Android by means of a malicious app that must be installed on the device, while another \cite{monkeysays} uses Google Firebase as a C\&C to self-issue voice commands to multiple smartphones. Differently from these works, \attackname{} is the first attack that uses self-activation against smart speakers to gain prolonged control over a dedicated VPA device, and the first work to evaluate self-issued hidden commands (see Appendix~\ref{attackwithhidden}). The main differences of \attackname{} and Lyexa \cite{lyexa} are: (i) \attackname{} does not make use of rogue speakers; (ii) \attackname{} can use different attack vectors; (iii) \attackname{} achieves persistence of the malicious skill on the device in a different way; (iv) Lyexa uses ultrasonic, inaudible voice commands; (v) \attackname{} does not need to dynamically append ``use Mask Attack'' to user commands, since the Mask Attack skill is started beforehand by the attacker. Table~\ref{comparisontable} summarizes the comparison among these solutions and \attackname{}.

\begin{table}[h!]
    \caption{Comparison Between \attackname{} and Other Attacks with Similar Strategy or Goal}
    \label{comparisontable}
    \begin{center}
    \resizebox{\columnwidth}{!}{
    \begin{tabular}{ |c|c|c|c|c|c|c|c|c|c| }
        \hline
        \multicolumn{3}{|c|}{\textbf{Attack Details}} & \multicolumn{3}{|c|}{\textbf{Does not rely on...}} & \multicolumn{4}{|c|}{\textbf{Features}} \\
        \hline
        \textbf{Paper} & \textbf{Target} & \textbf{Leverages} & \textbf{Soc. Eng.} & \textbf{2nd Speaker} & \textbf{Proximity} & \textbf{C\&C} & \textbf{HVC} & \textbf{Deceive} & \textbf{Persist} \\
        \hline
        Adv. Music \cite{adversarialmusic} & \faBullhorn & - & \greenyescheck & \rednocheck & \greenyescheck & - & - & - & - \\
        \hline
        Monkey \cite{monkeysays} & \faAndroid & SI & \rednocheck & \greenyescheck  & \greenyescheck  & \greenyescheck  & \rednocheck & - & \greenyescheck  \\
        \hline
        YVAIM \cite{yourvoiceassistantismine} & \faAndroid & SI & \rednocheck & \greenyescheck  & \greenyescheck  & \rednocheck & \rednocheck & - & \greenyescheck  \\
        \hline
        A11y \cite{a11yattacks} & \faAndroid, \faWindows & SI & \rednocheck & \greenyescheck  & \greenyescheck  & \rednocheck & \rednocheck & - & \greenyescheck  \\
        \hline
        Lyexa \cite{lyexa} & \faBullhorn & VMA & \greenyescheck  & \rednocheck & \rednocheck & \rednocheck & \greenyescheck  & \greenyescheck  & \greenyescheck  \\
        \hline
        \textbf{\attackname{} (BT)} & \faBullhorn & SI, VMA & \greenyescheck  & \greenyescheck  & \rednocheck & \rednocheck & \greenyescheck  & \greenyescheck  & \greenyescheck  \\
        \hline
        \textbf{\attackname{} (Radio)} & \faBullhorn & SI, SS, VMA & \rednocheck & \greenyescheck  & \greenyescheck  & \greenyescheck  & \greenyescheck  & \greenyescheck  & \greenyescheck  \\
        \hline
    \end{tabular}
    }
    \end{center}
    \footnotesize{\faAndroid: Android mobile phones. | \faWindows: Windows computers. | \faBullhorn: Smart speakers. | \textbf{SI:} Self-issue. | \textbf{SS:} Skill Squatting. | \textbf{Soc. Eng.:} Social Engineering, in this case \greenyescheck{} is assigned if the user does not have to inadvertently open malware (in form of applications, executables or skills), \rednocheck{} otherwise. | \textbf{Proximity:} \greenyescheck{} is assigned if the attack does not require the adversary (or a device controlled by them) to be near the target, \rednocheck{} otherwise. | \textbf{C\&C:} ability, for the attacker, to issue arbitrary commands to multiple compromised devices at once. | \textbf{HVC:} ability to use adversarial inputs or inaudible ones as payload during the attack. | \textbf{Deceive:} ability to interact with the user without being detected. We mark smartphone and computer attacks with ``-'' as they rely on malware that could hide the malicious behaviour. | 
    \textbf{Persist:} the attack keeps running unless uncommon circumstances occur.}
\end{table}

\section{Conclusion}
We introduced \attackname{}, an attack that plays audio tracks containing voice commands over an Echo device, which executes them as if issued by the legitimate user. We evaluated the performance of many self-issued commands, generated with different Google TTS voice profiles, and we presented the Full Volume Vulnerability (FVV) to enhance the reliability of self-issued commands. We observed that, when exploiting the FVV, an attacker can self-issue any command to an Echo device with a 99\% success rate in the small scenario, that is, when there are multiple obstacles close to Echo. In all other scenarios, we  observed an above-50\% success rate on average. We reported the vulnerabilities we found to Amazon, who rated them with Medium severity. \attackname{} does not require rogue speakers to be in proximity of the target device, and does not have heavy computational requirements, as TTS samples can be easily generated and stored for later use. Additionally, the exploitation flow is rather simple, as it is sufficient to connect Echo to one of the attack vectors. During the tests, we were able to successfully perform a set of malicious actions that an attacker could issue by leveraging \attackname{}, and we evaluated their feasibility and impact in a real scenario, by discussing the results of a field study and of a survey submitted to a study group composed of 18 Echo users.


\begin{acks}
Sergio Esposito's research was supported by a PhD studentship from Royal Holloway, University of London. Giampaolo Bella's research was supported by project MEGABIT "PIAno di inCEntivi per la RIcerca di Ateneo 2020/2022 (PIACERI)", University of Catania.
\end{acks}

\bibliographystyle{ACM-Reference-Format}
\footnotesize{\bibliography{ava}}
\normalsize

\appendix

\section{Analysis of Attack Vectors}\label{app:vectors}

\mypar{Attack Vector 1: Radio Station} To make the Echo device tune in a malicious radio station, the user must open a malicious skill first. This can be done in different ways: the attacker could trick the user into calling such malicious skill with a social engineering attack, or the same skill could be squatting another one, by means of homophones, compound words or phonetic confusion~\cite{skillsquatting}. Another way is giving the malicious skill an invocation name already used by another skill \cite{alexaisthisskillsafe}, so that it might be called instead of the legitimate one. Once tuned in, the radio station streams over Echo some voice commands, which the adversary had previously generated. When the Echo device captures a wakeword, the audio played by the radio station is turned down: this allows the device to hear the whole length of the self-issued voice command, satisfying the \textit{non-exclusivity} condition. Hence, this is a valid attack vector, and attackers can use it as a C\&C server to issue commands to multiple Echo devices at once.  Music and Radio skills allow developers to build a skill that plays a radio station, or a playlist of audio tracks. In this case, the radio station is preferable because its contents can be changed by the attacker in real time. However, Music and Radio skills are only available in US for the time being. Therefore, if not available, the adversary must choose a different attack vector.\\

\mypar{Attack Vector 2: Bluetooth Audio Streaming} If the adversary is in proximity of the Echo device they want to attack, they can use another device, such a smartphone or a computer, to connect via Bluetooth to the target Echo and make it act as a speaker for the adversary's device. This allows the attacker to play the audio files containing the malicious voice commands from the Echo's speaker. When the wakeword is captured, the volume of the streamed audio is turned down just like with Music and Radio skills -- hence the \textit{non-exclusivity} condition is met.

Using this attack vector, the adversary does not need to host the voice commands online or to have a publicly reachable malicious radio station, since they can store the voice commands on the Bluetooth device. Another advantage of this approach is that the attacker can leverage the Full Volume Vulnerability described in Section~\ref{exploitation} to increase the success rate of the self-issued commands. However, the adversary can only attack one Echo device at the time, and they need to be physically near the target for the Bluetooth connection to work properly.\\

\mypar{Attack Vector 3: SSML \texttt{audio} Tag} When coding a skill, the developer can use SSML to control how Alexa speaks. The \texttt{audio} SSML tag allows the skill developer to insert a mp3 file, as they might want to use their own voice instead of Alexa's one, or play a short song at some point. However, if Echo hears a wakeword while reproducing an \texttt{audio} tag, such audio is paused instead of being turned down. This would interrupt the self-issue of any voice command: hence, the \textit{non-exclusivity} condition is not met and \texttt{audio} tags cannot be used.\footnote{At times, there can be a short delay between the wakeword being said and its recognition by the Echo device. This could allow a very short command, such as ``Echo, hi'' to be successfully self-issued with an \texttt{audio} tag.} Due to this, we consider only Radio skills and Bluetooth streaming as valid attack vectors.

\section{Setup of the Experiments}
\begin{figure}[!hbt]
    \centering
    \begin{minipage}[b]{0.15\textwidth}
        \includegraphics[width=\textwidth]{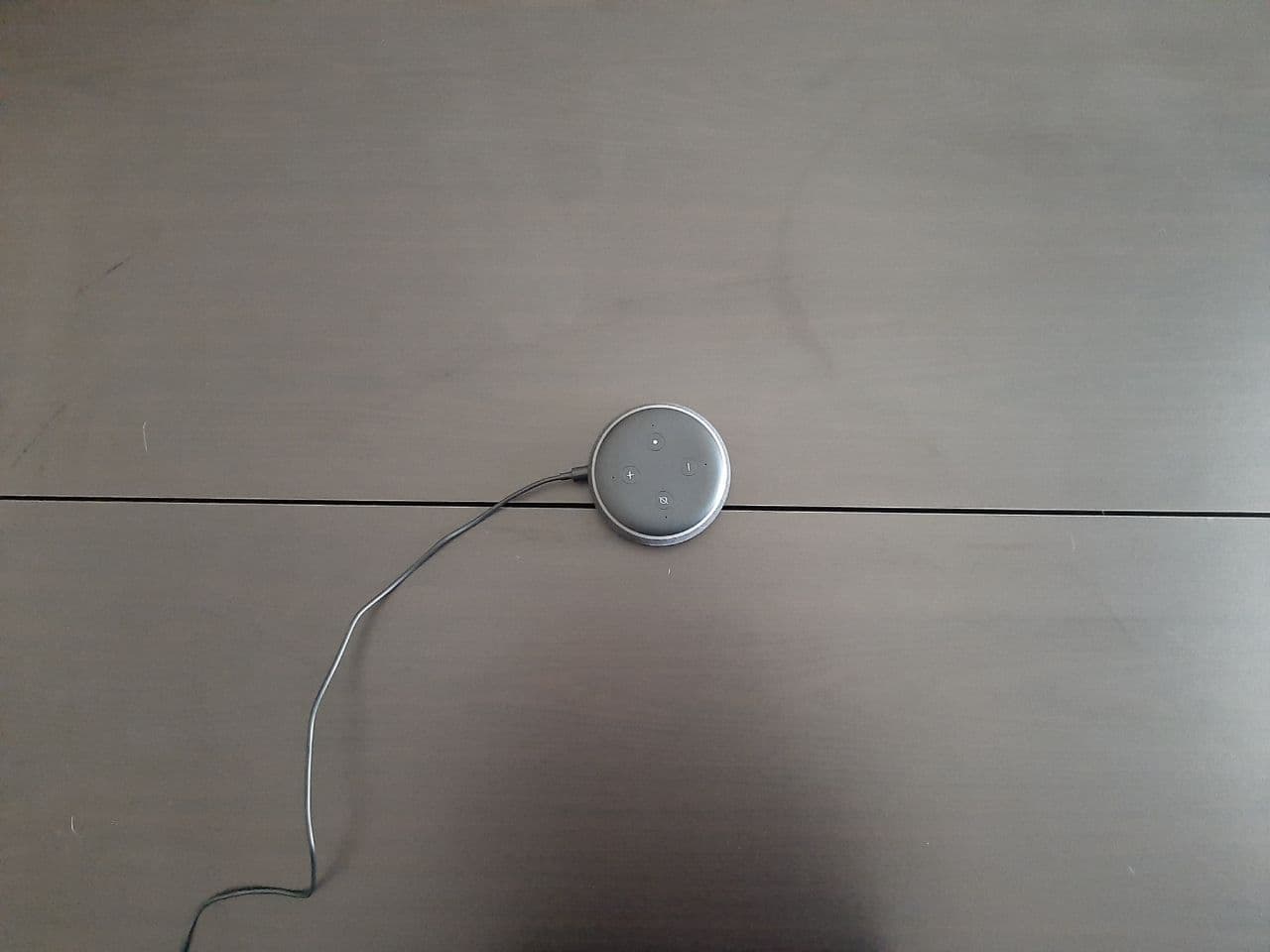}
    \end{minipage}
    \begin{minipage}[b]{0.15\textwidth}
        \includegraphics[width=\textwidth]{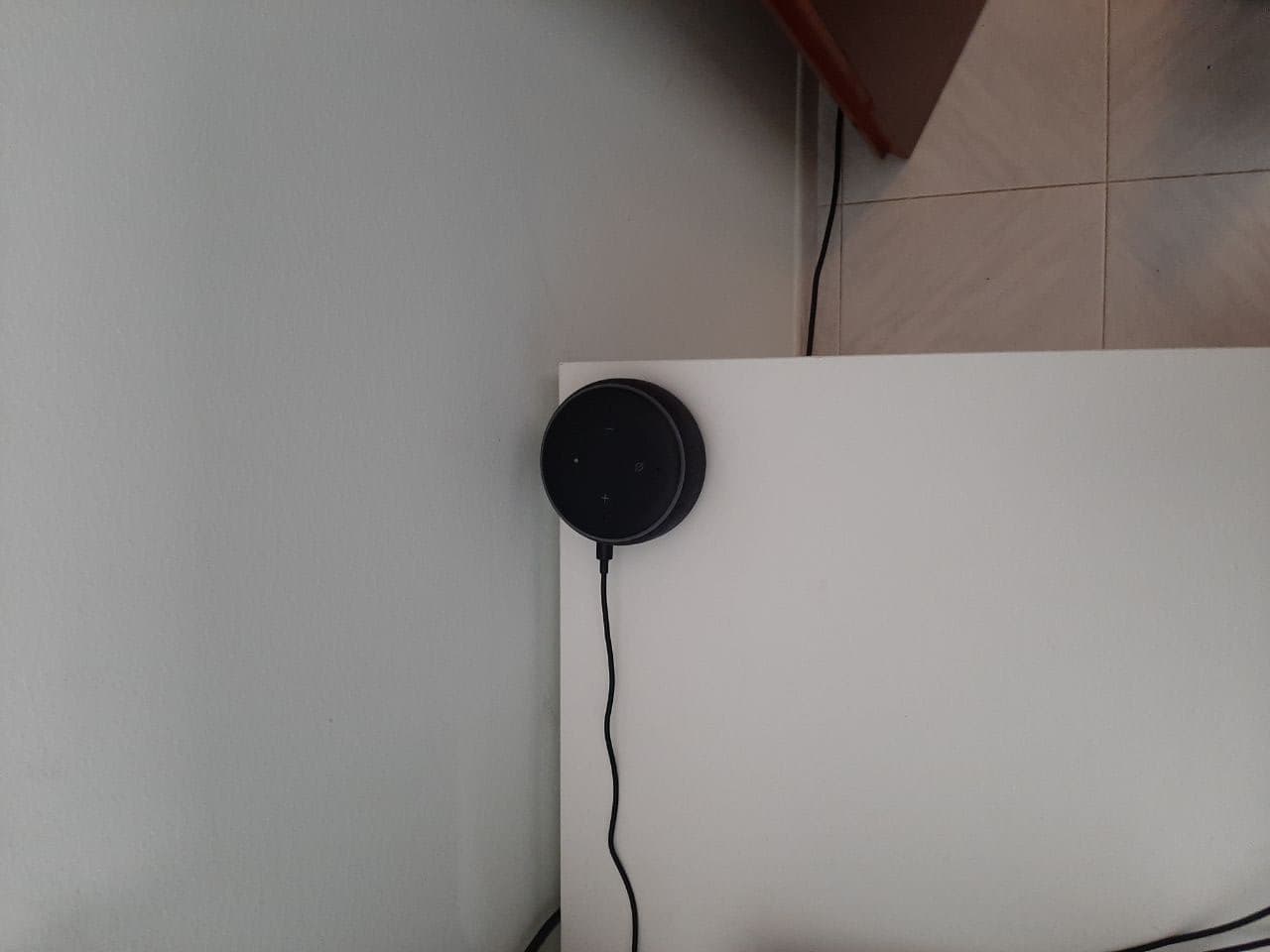}
    \end{minipage}
    \begin{minipage}[b]{0.15\textwidth}
        \includegraphics[width=\textwidth]{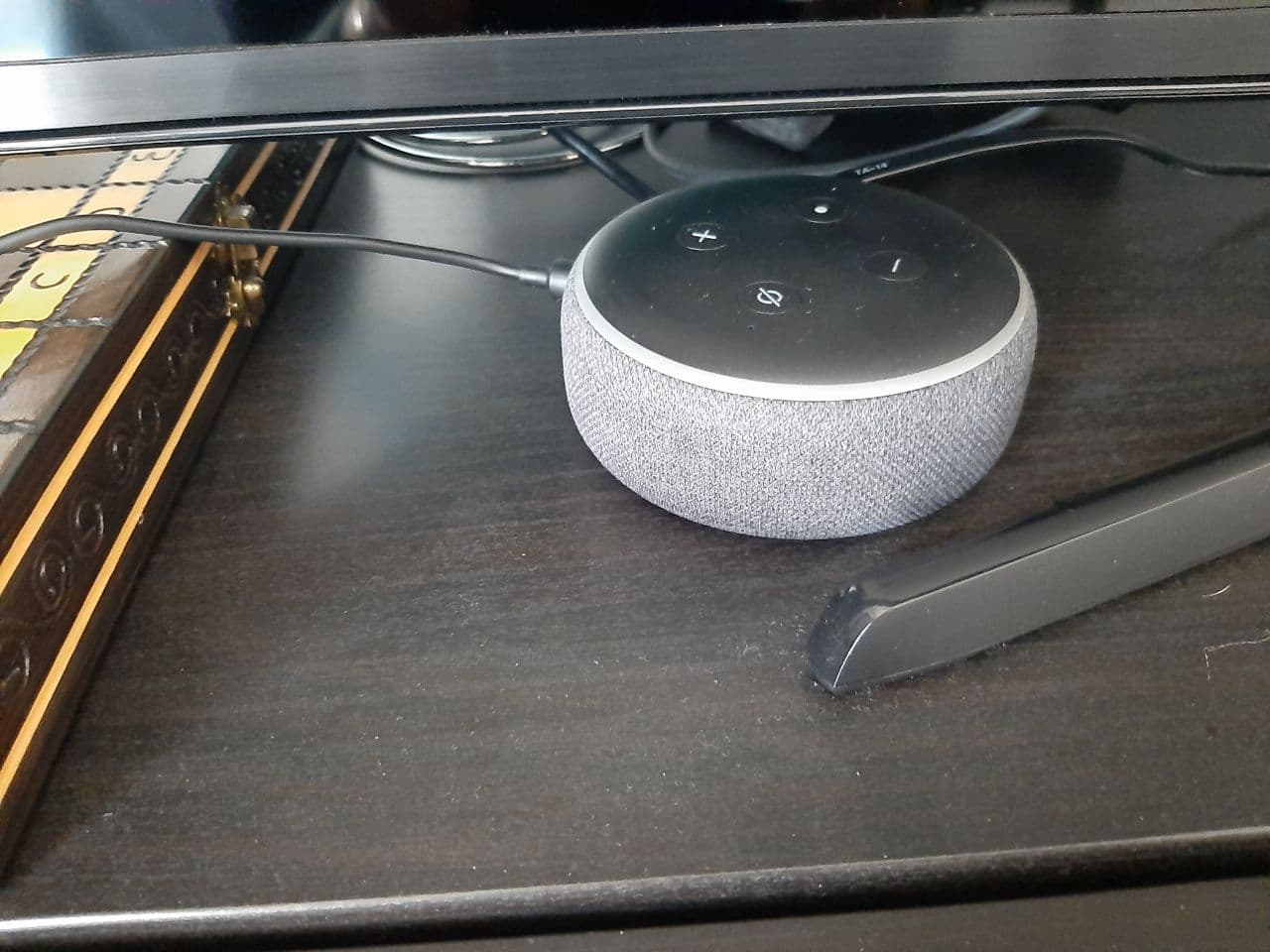}
        
    \end{minipage}
    \caption{Placement of the Echo Device in the Open (left), Wall (center) and Small (right) Scenarios.}
    \label{fig:placements}
\end{figure}

Figure~\ref{fig:placements} shows typical placement of the 3rd Generation Echo Dot during our experiments for the Open, Wall and Small scenarios.

\section{Adversarial Payload Performance}
\label{attackwithhidden}
Adversarial samples were generated with Devil's Whisper DockerHub~\cite{dockerhubdw} and tested using different parameters to measure the best performing combination. In particular, we verified that the two key parameters to improve the success rate of the hidden commands are \texttt{mini\_noise\_value} and \texttt{aspire\_noise\_value}, which alter the Signal-To-Noise Ratio (SNR) for the adversarial samples. We noted that, when using the default value of 5,000 for both variables, the Echo device could not be activated when the samples were self-issued, and only a small number (3\%) succeeded in activating it when played in its close proximity. Increasing the noise value to 5,500, we observed a dramatic improvement of the activation rate when the samples were not self-issued: in fact, 83.5\% of the samples were successful. We noticed that the first self-activation of the Echo device happened at noise 7,500, and the first complete self-issue (both wakeword and command) of an adversarial input was at noise 8,000. We also verified that, by increasing the noise values past 11,500, the success rate of the samples decreased noticeably, due to clipping or excessive distortion of the audio.

In the tests, \attackname{} was able to successfully self-activate the Echo device with adversarial samples in the small space scenario. The whole command could be recognized by exploiting the FVV (\S\ref{attackwithplain}) or, alternatively, \attackname{} exploited the adversarial samples to refresh the Mask Attack skill timer in a stealthy fashion. In fact, as noted in Section~\ref{persistency}, only the wakeword needs to be recognized to achieve this goal. Table~\ref{hvctable} reports the percentage of the adversarial \texttt{mini} samples that triggered at least one self-activation of the Echo device or one self-execution of the command ``Echo, turn off the light'', over 100 commands per each noise value. It can be observed that the adversarial samples generated on top of the songs labeled \textit{Song 1} and \textit{Song 3} are the most effective in self-issuing complete commands. When these tracks are played again to check their reliability, we find that they replicate this behaviour 15\% of the times on average. 

\begin{table}[h!]
    \caption{Evaluation of Adversarial Samples in the Small Space Scenario with Varying Noise and Background Songs}
    \label{hvctable}
    \begin{center}
    \begin{tabular}{ |c|c c|c c|c c|c c| }
        \hline
        \multirow{3}{*}{Track} & \multicolumn{8}{|c|}{Noise Value and Behaviour} \\ \cline{2-9}
        & \multicolumn{2}{|c|}{8000} & \multicolumn{2}{|c|}{9000} & \multicolumn{2}{|c|}{10000} & \multicolumn{2}{|c|}{11000} \\ \cline{2-9}
        & Ac & Ex & Ac & Ex & Ac & Ex & Ac & Ex \\
        \hline
        Song 1 & 3\% & 1\% & 16\% & 2\% & 15\% & 2\% & 13\% & 1\% \\
        Song 2 & 14\% & 0\% & 13\% & 0\% & 12\% & 0\% & 2\% & 0\% \\
        Song 3 & 0\% & 0\% & 0\% & 0\% & 3\% & 1\% & 8\% & 1\% \\
        Song 4 & 0\% & 0\% & 0\% & 0\% & 0\% & 0\% & 0\% & 0\% \\
        Song 5 & 1\% & 0\% & 1\% & 0\% & 0\% & 0\% & 0\% & 0\% \\
        \hline
    \end{tabular}
    \end{center}
    \footnotesize{\textbf{Ac:} \% of \texttt{mini} samples that triggered self-activation of the Echo device \\ \textbf{Ex:} \% of \texttt{mini} samples that were successfully self-executed}
\end{table}

\section{User Survey}
\label{app:survey}

Table~\ref{table:questions} lists all questions answered by the study group. Questions are divided in three categories. \emph{Usage Information} is related on how the users interact with their device. This information was used to determine the overall confidence of the user with the device and to understand the quality of their answers. \emph{Scenario Recognition} questions are needed to collect information on the different placements and volumes of real Echo devices, to understand the distribution of the different scenarios. \emph{Limitations Assessment} questions allow us to estimate the likelihood of the limitations to actually take place in real scenarios. To avoid bias, questions were presented to the users with no categorization and with uniform IDs (e.g., Q1, Q2, Q3). All participants who were involved in the field study also answered the survey. Out of all 18 participants, four people were aged between 18 and 24 years, eleven between 25 and 31 years, while the remaining three were aged 32 years or more. Thirteen participants identified themselves as male, and five as female.

\begin{table}[h]
    \caption{Survey for the Study Group}
    \label{table:questions}
    \begin{center}
    \begin{tabular}{ |c|p{22em}| }
        \hline
        \textbf{ID} & \textbf{Question} \\
        \hline
        \hline
        \rowcolor{gray!25}
        \multicolumn{2}{|l|}{\textbf{Usage Information}} \\
        \hline
        U1 & From 1 to 10, how would you score your skill in interacting with Amazon Echo and Alexa? \\
        \hline
        U2 & How often do you use your Echo device (e.g., once a day)? \\
        \hline
        U3 & What do you usually use your Echo device for? If you want, you can answer with some of the most common commands you use. \\
        \hline
        \rowcolor{gray!25}
        \multicolumn{2}{|l|}{\textbf{Scenario Recognition}} \\
        \hline
        S1 & Can you send a photo that shows the placement of your Echo device? Please make sure that the photo does not show people or personal data. If this is not possible, can you estimate the distance (in cm) of your Echo device from the nearby walls and obstacles? \\
        \hline
        S2 & What is the average volume of your Echo device? If you do not know, the current volume will be fine (ask ``Alexa, what is the current volume?''). \\
        \hline
        \rowcolor{gray!25}
        \multicolumn{2}{|l|}{\textbf{Limitations Assessment}} \\
        \hline
        L1 & Have you ever unplugged your Echo device from the power source? Why? \\
        \hline
        L2 & Have you ever connected headphones to your Echo device? How often do you do so? \\ 
        \hline
        L3 & How do you usually change Alexa's volume? \\
        \hline
        L4 & How often do you turn off Echo's microphone? \\
        \hline
        L5 & Do you know what the green light ring around Echo means? If you do not know and want to learn what it means, please search on the Internet AFTER answering this question. \\
        \hline
        L6 & Have you ever used Echo as a Bluetooth speaker (i.e. connecting another device via Bluetooth to play music or other audio files)? \\
        \hline
        L7 & Do you follow any security best-practices (e.g., you turn off Echo's microphone after 9pm)? \\
        \hline
    \end{tabular}
    \end{center}
\end{table}

Table~\ref{table:survey} summarizes the results of the survey, where green colour indicates results that are favourable for \attackname{}, red unfavourable scenarios and orange mostly neutral results. As we can see, the average volume of the devices is set at 4.7, which is ideal for \attackname{}'s success rate, given that \attackname{} commands at volume 4 and 5 perform well (\S\ref{evaluation}). Additionally, most of the users place their Echo device in the Small Space scenario, which allows \attackname{} to achieve the best performance (\S\ref{evaluation}).
Only 27\% of users ever rebooted their device voluntarily, mostly to move the device to another room, or for cleaning purposes. Only 11\% of the users ever switched off their Echo due to a malfunctioning or to perform a hard reboot, and only 6\% turned off the device because they did not recognize the color of the light ring. This is favourable for \attackname{}, because restarting the device is a way to disconnect Echo from an attack vector. No one ever connected headphones to Echo -- their connection would not allow the self-issued commands to be captured by Echo's microphone. However, because 0\% of users ever used them, we can assume this behaviour is very unusual among Echo users, and this limitation remains theoretical.
Only 27\% of participants knew that the green light indicates an ongoing call, hence the remaining 73\% would not realize that there is an ongoing rogue call. Only 11\% of users claimed to mute Echo's microphone sometimes -- all the other participants reported they do it rarely, very rarely or never. Hence, the Echo device will keep listening at self-issued commands in most cases. Finally, only 6\% of users reported to systematically follow security procedures, such as turning off Echo's microphone during the night -- this means that most devices would be vulnerable during night time or when the user is away. Hence, the results of this survey clearly show that most of the limitations do not impact the feasibility and success rate of \attackname{}. 

\begin{table}[hbt!]
    \caption{Survey Results}
    \label{table:survey}
    \footnotesize
    \begin{centering}
    \begin{tabular}{ l l l }
        
        \Xhline{2\arrayrulewidth}
        \multicolumn{2}{ l }{\textbf{Test}} & \textbf{Result} \\
        \hline
        
        \multirow{3}{*}{Device Placement} & Open Scenario & \textcolor{darkgreen}{16.67\%} \\
        & Wall Scenario & \textcolor{orange}{27.78\%} \\
        & Small Scenario & \textcolor{darkgreen}{55.55\%} \\
        
        \multicolumn{2}{ l }{Average Volume} & \textcolor{darkgreen}{4.7} \\
        
        \multicolumn{2}{ l }{Users who voluntarily ever switched off Echo} & \textcolor{orange}{27\%} \\
        
        \multicolumn{2}{ l }{Users who systematically switch off Echo} & \textcolor{darkgreen}{6\%} \\
        
        \multicolumn{2}{ l }{Users who ever connected headphones} & \textcolor{darkgreen}{0\%} \\
        
        \multicolumn{2}{ l }{Users who change volume via manual commands*} & \textcolor{orange}{27\%} \\
        
        \multirow{3}{*}{Users who turn off Echo's microphone} & Never & \textcolor{darkgreen}{66.67\%} \\
        & Rarely / Very Rarely & \textcolor{darkgreen}{22.22\%} \\
        & Sometimes & \textcolor{darkgreen}{11.11\%} \\
        
        \multicolumn{2}{ l }{Users who use Echo as a Bluetooth speaker} & \textcolor{orange}{44.44\%} \\
        
        \multicolumn{2}{ l }{Users who know the meaning of the green light ring} & \textcolor{orange}{27\%} \\
        
        \multicolumn{2}{ l }{Users who follow security procedures} & \textcolor{darkgreen}{6\%} \\
        
        \Xhline{2\arrayrulewidth}

    \end{tabular}
    \end{centering}\\
    \textbf{*} Including those who stated to use both voice and manual commands.
\end{table}

\section{Ethics and Responsible Disclosure}
\label{responsibledisclosure}

\mypar{Ethics of the Tests}
Regarding the field study, we requested all users in the three households permission to place our Echo device in a specific room, and to collect usage data useful for the experiment, such as the given commands or the actions they performed during the different tests, and to use these data in an anonymised form only for the purpose of this research work.
After the tests were performed, we explained the users the tests we did, and informed them of the results: all the users were again asked to either confirm or withdraw their permissions to use the results in a scientific study. At no points were any of the user's private data accessed by third parties or by the authors of this study. In addition, all user utterances saved on the Mask Attack database and on the Amazon Cloud (including recordings of the commands that are stored on the Amazon Cloud by default) were deleted after summarizing in an anonymised format the results of the tests. Note that all the tests performed on Echo only involved accounts, emails and phone numbers owned solely by the authors of this work, and no user nor third-party data were accessed during the experiments. Regarding the user survey, all the questionnaires were administered online through secure means, and users were asked for their explicit consent to use the collected information, in an anonymised format, for a research work. Participation to the survey was voluntary, and users were informed they could withdraw their consent at any time. Finally, the permission forms of the field study and of the survey were extensively analyzed through an internal formal assessment, as mandated by our institutional policy, and these activities did not identify any danger or ethical concern.

\vfill\eject
\mypar{Responsible Disclosure}
We dutifully contacted Amazon via their Vulnerability Research Program (VRP), reporting the existence of all three vulnerabilities we have found, that is, the self-issue vulnerability, the Full Volume Vulnerability and the SSML \texttt{break} tag chain policy violation. Our report was submitted on January 21\textsuperscript{st}, 2021, and the Amazon VRP Team quickly reacted to it, asking for more details. On February 18\textsuperscript{th}, 2021, our research team engaged in a call with the Amazon VRP Team to discuss details of the attack, its impact and possible mitigation strategies. Our report was rated with Medium severity by the Amazon VRP team. We have also provided a draft copy of this manuscript to Amazon VRP Team for knowledge. To date (November 2021), all three vulnerabilities are pending resolution or mitigation, although Amazon VRP team has confirmed that they are working on some fixes that are targeted for December 2021: the Amazon VRP Team did not object to us submitting this final research work. We have not disclosed any detail of the attack or any part of the related source code to any third party. In addition, in this work we have identified possible countermeasures.

\end{document}